\documentclass[review]{elsarticle}

\usepackage{lineno,hyperref}
\usepackage{bm, amssymb, mathtools, pifont}
\usepackage{longtable}
\usepackage{tabularx}
\usepackage{gensymb}
\usepackage{amsmath}
\usepackage{multirow}
\usepackage{comment}
\usepackage{graphicx}
\usepackage{booktabs}
\usepackage{subfig}			
\usepackage{caption}
\usepackage{gensymb}

\usepackage{footnote}
\makesavenoteenv{table}

\modulolinenumbers[5]

\journal{Acta Astronautica}

\begin{document}

\begin{frontmatter}

\title{Adaptive Generalized ZEM-ZEV Feedback Guidance for Planetary Landing via a Deep Reinforcement Learning Approach}

\author{Roberto Furfaro\fnref{Roberto}\corref{mycorrespondingauthor}}
\cortext[mycorrespondingauthor]{Corresponding author: Roberto Furfaro}
\address{Department of Systems \& Industrial Engineering, Department of Aerospace and Mechanical Engineering, University of Arizona, Tucson, AZ 85721, USA}
\ead{robertof@email.arizona.edu}
\fntext[Roberto]{Professor, Department of System \& Industrial Engineering, Department of Aerospace and Mechanical Engineering, University of Arizona, Tucson, AZ 85721, USA}

\author{Andrea Scorsoglio\fnref{Andrea}}
\address{Department of Systems \& Industrial Engineering, University of Arizona, Tucson, AZ 85721, USA}
\ead{andreascorsoglio@email.arizona.edu}
\fntext[Andrea]{PhD student, Department of System \& Industrial Engineering, University of Arizona, Tucson, AZ 85721, USA}

\author{Richard Linares\fnref{Richard}}
\address{Department of Aeronautics and Astronautics, Massachusetts Institute of Technology, Cambridge, MA, 02139, USA}
\ead{linaresr@mit.edu}
\fntext[Richard]{Charles Stark Draper Assistant Professor, Department of Aeronautics and Astronautics, Massachusetts Institute of Technology, Cambridge, MA, 02139, USA}

\author{Mauro Massari\fnref{Mauro}}
\address{Department of Aerospace Science and Technology, Politecnico di Milano, Milan, 20156, ITA}
\ead{mauro.massari@polimi.it}
\fntext[Mauro]{Associate Professor, Department of Aerospace Science and Technology, Politecnico di Milano, Milan, 20156, ITA}


\begin{abstract}
    Precision landing on large and small planetary bodies is a technology of utmost importance for future human and robotic exploration of the solar system. In this context, the Zero-Effort-Miss/Zero-Effort-Velocity (ZEM/ZEV) feedback guidance algorithm has been studied extensively and is still a field of active research. The algorithm, although powerful in terms of accuracy and ease of implementation, has some limitations. Therefore with this paper we present an adaptive guidance algorithm based on classical ZEM/ZEV in which machine learning is used to overcome its limitations and create a closed loop guidance algorithm that is sufficiently lightweight to be implemented on board spacecraft and flexible enough to be able to adapt to the given constraint scenario. The adopted methodology is an actor-critic reinforcement learning algorithm that learns the parameters of the above-mentioned guidance architecture according to the given problem constraints.
\end{abstract}

\begin{keyword}
    Optimal Landing Guidance, Deep Reinfocement Learning, Closed-loop Guidance
    \MSC[2019] 00-01\sep  99-00
\end{keyword}

\end{frontmatter}

\linenumbers

\section{Introduction}

Precision landing on large and small planetary bodies is a technology of utmost importance for future human and robotic exploration of the solar system. 
Over the past two decades, landing systems for robotic Mars missions have been developed and successfully deployed robotic assets on the Martian surface (e.g. rovers, landers)\cite{shotwell2005phoenix,grotzinger2012mars}. Considering the strong interest in sending humans to Mars within the next few decades, as well as the renewed interest in building infrastructure in the Earth-Moon system for easy access to the Lunar surface \cite{burns2018science}, the landing system technology will need to progress to satisfy the  demand for more stringent requirements. The latter will call for guidance systems capable of delivering landers and/or rovers to the selected planetary surface with higher degree of precision. In the case of robotic Mars landing, the 3-sigma ellipse, which describes the landing accuracy, has seen a dramatic improvement from the 100 km\cite{shotwell2005phoenix} required by the Phoenix mission to 5 km featured by the newly developed "Sky Crane" system which delivered the Mars Science Laboratory (MSL) to the martian surface in 2012\cite{steltzner2010mars}. Although such improvements were needed to deliver a better science through robotic devices, future missions may require delivering cargo (including humans) in specified location with pinpoint accuracy (somewhere between 10 and 100 meters). Importantly, delivering scientific packages in geologically interesting locations may require guidance systems that are fuel-optimal while satisfying stringent flight constraints (e.g. do not crash on the surfaces with elevated slope).

One of the most important enabling technology for planetary landing is the powered descent guidance algorithm. Generally, powered descent indicates a phase in the landing concept of operation where rockets provide the necessary thrust to steer the spacecraft trajectory toward the desired location on the planetary surface. The corresponding guidance algorithm must determine in real-time both thrust magnitude, directions and time of flight. The original Apollo guidance algorithm, which was used to drive the Lunar Exploration Module (LEM) to the lunar surface, was based on an iterative approach that computed  on the ground a flyable reference trajectory in the form of a quartic polynomial\cite{klumpp1974apollo}. The real-time guidance algorithm generated an acceleration command that targets the final condition of the trajectory. A variation of the Apollo guidance was also employed for the MSL powered descent phase \cite{singh2007guidance}.
Over the past two decades, there has been a tremendous interest in developing new classes of guidance algorithms for powered descent that improve performance over the classical Apollo algorithm both in precision and fuel-efficiency. Trajectory optimization methods are currently playing a major role in generating feasible, fuel-efficient trajectories that can be potentially computed in real-time. Much effort has been placed in transforming a fuel-optimal constrained landing problem in a convex optimization problem that can guarantee finding the global optimal solution in a polynomial time \cite{acikmese2007convex,blackmore2010minimum}. Such approach yielded the G-FOLD algorithm \cite{trawny2015flight} which has been recently tested in real landing systems. Importantly, the convexification methodology has been recently applied to other aerospace guidance problems. A review of the application areas can be found in \cite{liu2017survey}. Conversely, another class of popular methods generally employed to solve optimal guidance problems, rely on the application of Pontryagin Minimum Principle (PMP). Named indirect methods, such algorithms solve the  Two-Point Boundary Value Problems (TPBVP) arising from the necessary conditions for optimality. Recently,  a three-dimensional, fuel-optimal, powered descent guidance algorithm based on indirect methods has been developed \cite{lu2017propellant}. The approach, generally named Universal Powered Guidance (UPG), relies on a general powered descent methodology which has been developed and applied to ascent and orbital transfer problems by Ping Lu over the past decade \cite{lu2010highly,lu2008rapid,lu2012versatile}. The algorithm is capable of delivering both human and robotic device on planetary surfaces efficiently and accurately\cite{lu2018adaptive}. UPG provides a robust approach based on indirect methods to 1) analyze the thrust profile structure (i.e. analyze the bang-bang profile)and 2) find the optimal numbers of burn times. Importantly, the advantage over G-FOLD is due to its simplicity and flexibility as it does not require customization of the algorithm\cite{lu2017propellant}. However, UPG has the disadvantage that both inequality and thrust direction constraints are generally difficult to enforce\cite{lu2017propellant}.

Besides the above mentioned methods, over the past few years, researchers have been exploring the performances of the generalized Zero-Effort-Miss/Zero-Effort-Velocity (ZEM/ZEV) feedback guidance algorithm \cite{Guo:2013:ZMZV_generalized,Guo:2011:ZMZV_plan_landing_asteroid_inter} in the context of landing on large and small bodies of the solar system. The feedback ZEM/ZEV guidance law is analytical in nature and derived by a straightforward application of the optimal control theory to the power descent landing problem. The algorithm generates a closed-loop acceleration command that minimizes the overall system energy (i.e. the integral of the square of the acceleration norm).  The ZEM/ZEV feedback guidance is attractive because of its analytical simplicity and accuracy: guidance mechanization is straightforward and it can theoretically drive the spacecraft to a target location on the planetary surface both autonomously and with minimal guidance errors. Moreover, it has been shown to be globally finite time stable and robust to uncertainties in the model if a proper sliding parameter is added (Optimal Sliding Guidance)\cite{Furfaro:2016:Robust_ZEMZEV}. Although attractive because of its simplicity and analytical structure, the algorithm is not generally capable of enforcing either thrust constraints and/or flight constraints. There have been attempts to incorporate constraints in the classical ZEM/ZEV algorithm with the utilization of intermediate waypoints \cite{guo2013waypoint,Furfaro:2017:waypoints}. Although they report good performances, they lack of flexibility and ability to adapt in real-time.

In this paper, we propose a ZEM/ZEV-based guidance algorithm for powered descent landing that can adaptively change both guidance gains and time-to-go to generate a class of closed-loop trajectories that 1) are quasi-optimal (w.r.t. the fuel-efficiency) and 2) satisfy flight constraints (e.g. thrust constraints, glide slope). The proposed algorithm exploit recent advancements in deep reinforcement learning (e.g. deterministic policy gradient \cite{Silver:2014:deterministic_PG}), and machine learning (e.g. Extreme Learning Machines, ELM \cite{Huang:2015:ELM,Huang:2011:ELM_survey}).  
The overall structure of the guidance algorithm is unchanged with respect to the classical ZEM/ZEV, but the optimal guidance gains are determined at each time step as function of the state via a parametrized learned policy. This is achieved using a deep reinforcement learning method based on an actor-critic algorithm that learns the optimal policy parameters minimizing a specific cost function. The policy is stochastic, but only its mean, expressed as a linear combination of radial basis functions, is updated by stochastic gradient descent. The variance of the policy is kept constant and is used to ensure exploration of the state space. The critic is an Extreme Learning Machine (ELM) that approximates the value function. The approximated value function is then used by the actor to update the policy. The power of the method resides in its capability, if an adequate cost function is introduced, of satisfying virtually any constraint and in its model-free nature that, given an accurate enough dynamics simulator for the generation of sample trajectories, allows learning of the guidance law in any environment, regardless of its properties. This greatly expands the capabilities of classical ZEM/ZEV guidance, allowing for its use in a wide variety of environment and constraint combinations, giving results that are generally close to the constrained fuel optimal off-line solution. Additionally, because the guidance structure is left virtually unchanged, we are able to ensure that the adaptive algorithm is maintained globally stable regardeless of the gain adaptation.

The paper is organized as follows. In section \ref{sec:setup} the landing problem set-up is described. In section \ref{sec:theoretical_back}, the theoretical background is provided, including derivation of the classical ZEM/ZEV algorithm, deep Actor-Critic method and ELM. In section \ref{sec:adaptive_ZEM_ZEV}, the proposed adaptive ZEM/ZEV algorithm is described. In section \ref{sec:results}, numerical results for Mars landing scenarios are reported. In section \ref{chap:stability}, a stability analysis is conducted to show the global stability properties of the proposed algorithm. Conclusions are reported in section \ref{sec:conclusion}.

\section{Problem setup} \label{sec:setup}
\begin{figure}
    \centering
    \includegraphics[width=.90\textwidth]{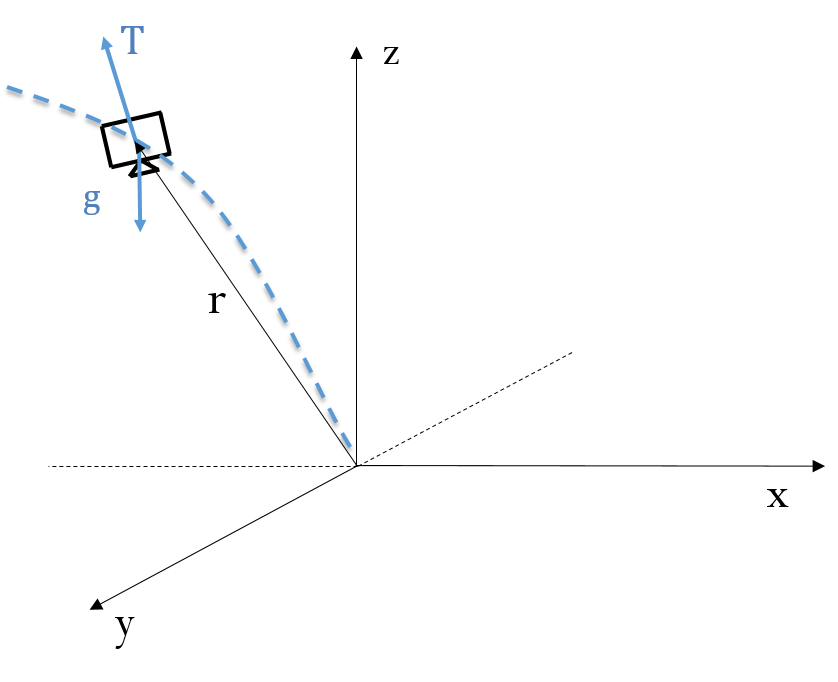}
    \caption{Problem setup}
    \label{fig:setup}
\end{figure}

The algorithm is being developed for a Mars soft landing scenario described in Figure \ref{fig:setup}. 
The problem is described in a orthonormal reference system centered on the nominal landing target on the ground. The equations of motion governing the dynamics of the problem expressed in the above mentioned reference frame are:
\begin{align}
& \mathbf{\ddot{x}}=\mathbf{g(\mathbf{r})}+ \dfrac{\mathbf{T}}{m} \\
& \dot{m}=-\dfrac{\Vert\mathbf{T}\Vert}{I_{sp}g_0}
\end{align}
where $\mathbf{x}=\left[\mathbf{r} , \mathbf{v} \right]^{T}$ is the state, $\mathbf{g(\mathbf{r})}$ is the gravity vector at position $\mathbf{r}$ and $\mathbf{T}$ is the thrust vector:
\begin{align}
    \mathbf{T} &= \begin{bmatrix}
           T_{x} \, , \,
           T_{y} \, , \,
           T_{z}
         \end{bmatrix}^T
\end{align}
It should be noted that the control policy is based on Zero-Effort-Miss/Zero-Effort-Velocity guidance which outputs an acceleration command rather than a thrust command. The thrust $\mathbf{T}$ is recovered indirectly knowing engine specifications and mass. The gravitational acceleration is aligned with the vertical direction at all times. The rotation of the planet is neglected as we consider only the terminal guidance of the power descent phase for a pinpoint landing problem where the altitude is small with respect to the radius of the planet. The interaction with the thin martian atmosphere is also neglected. The spacecraft is constrained to remain above the ground which has a constant slope angle of 4\degree  with respect to the horizontal direction except for 5 meter radius flat area around the target.


\section{Theoretical Background}
\label{sec:theoretical_back}


\subsection{Classical Zero-Effort-Miss/Zero-Effort-Velocity algorithm}
\label{sec:Classical_ZEMZEV}

Consider the problem of landing on a large planetary body of interest with mission from time $t_0$ to $t_f$. The unconstrained, energy-optimal guidance problem can be formulated as follows: Find the overall  acceleration \textbf{a}  that minimizes the performance index:
\begin{equation}
J=\dfrac{1}{2} \int_{t_0}^{t_f} \mathbf{a^T} \mathbf{a} \quad \text{dt}
\end{equation}
for a spacecraft subjected to the following general dynamic equations, valid in any case, even for non-inertial systems:
\begin{align} \label{eq:eom}
\begin{split}
& \dot{\mathbf{r}}=\mathbf{v}\\
& \dot{\mathbf{v}}=\mathbf{a}+\mathbf{g}\\
& \mathbf{a}=\mathbf{T}/m
\end{split}
\end{align}
with $\mathbf{r}$, $\mathbf{v}$, $\mathbf{T}$ and \textbf{a} being position, velocity, thrust and acceleration command vectors respectively and $\mathbf{g}$ is the gravitational acceleration. In the remainder of the paper,  $\mathbf{g}$ is assumed to be constant. The latter works well for modeling the powered descent guidance starting close to the planetary surface of a large body. Additionally, note that additional forces (e.g. aerodynamics forces experienced by bodies close to the Mars surface) are considered negligible. The following boundary conditions are given:
\begin{equation}
\mathbf{r}(t_0)=\mathbf{r_0}, \qquad \mathbf{r}(t_f)=\mathbf{r_f}
\end{equation}
\begin{equation}
\mathbf{v}(t_0)=\mathbf{v_0}, \qquad \mathbf{v}(t_f)=\mathbf{v_f}
\end{equation}
Importantly,  no constraints on acceleration and on the spacecraft state are assumed. The necessary conditions can be derived by a straightforward application of the PMP. Indeed, the Hamiltonian function for this problem is then defined as 
\begin{equation}
\mathbf{H}=\dfrac{1}{2} \mathbf{a^T} \mathbf{a} + \mathbf{p_r}^T \mathbf{v} + \mathbf{p_v}^T (\mathbf{g}+\mathbf{a})
\end{equation}
where $\mathbf{p_r}$ and $\mathbf{p_v}$ are the costate vectors associated with position and velocity vector respectively.
The time-to-go is defined as: $t_{go}=t_f-t$. The optimal acceleration at any time $t$, can be found by directly applying the optimality condition as
\begin{equation} \label{eq:acc1}
\mathbf{a}=-t_{go}\mathbf{p_r}(t_f)-\mathbf{p_v}(t_f)
\end{equation}
By substituting equation \ref{eq:acc1} into the dynamics equations to solve for $\mathbf{p_r}(t_f)$ and $\mathbf{p_v}(t_f)$, the optimal control solution with specified $\textbf{r}_f$ and $\textbf{v}_f$ and $t_{go}$ is obtained as:
\begin{equation} 
\label{eq:optZEMZEV1}
\mathbf{a}=\dfrac{6[\mathbf{r_f} - (\mathbf{r}+t_{go}\mathbf{v})]}{t_{go}^2} - \dfrac{2(\mathbf{v_f}-\mathbf{v})}{t_{go}} + 
\dfrac{6 \int_{t}^{t_f}(\tau-t)\mathbf{g}d\tau}{t_{go}^2} - \dfrac{4\int_{t}^{t_f}\mathbf{g}d\tau}{t_{go}}
\end{equation}
The Zero-Effort-Miss (ZEM) and the Zero-Effort-Velocity (ZEV) are defined, respectively, as the distance between the desired final position and velocity and the projected final position and velocity if no additional control is commanded from time $t$ onward. Consequently, ZEM and ZEV have the following expressions:
\begin{align}
\label{eq:ZEM_ZEV_gt_def}
\begin{split}
& \mathbf{ZEM}=\mathbf{r}_f- \left[ \mathbf{r}+t_{go}\mathbf{v}+\int_{t}^{t_f}(t_f-\tau)\mathbf{g}(\tau)d\tau \right]\\
& \mathbf{ZEV}=\mathbf{v}_f-\left[ \mathbf{v} + \int_{t}^{t_f}\mathbf{g}(\tau)d\tau \right]
\end{split}
\end{align}
Then the optimal control law \ref{eq:optZEMZEV1} can be expressed as:
\begin{equation}
\label{eq:classical_ZEM_ZEV}
\mathbf{a}=\dfrac{6}{t_{go}^2}\mathbf{ZEM} - \dfrac{2}{t_{go}}\mathbf{ZEV}
\end{equation}
Note that the solution holds also in the case where $\mathbf{g} = \mathbf{g}(t)$. In any other case in which $\mathbf{g}$ is neither constant nor time dependant, the control law is still usable but it will not be necessarily optimal. In case the equations of motion are non-linear and in general when \ref{eq:ZEM_ZEV_gt_def} do not apply, ZEM and ZEV are expressed in a slightly different way. The projected position and velocity cannot be recovered analytically: they must be obtained through an integration of the equations of motion from the current time instant to the end of the mission with control actions set to zero.
\begin{align}
\label{eq:ZEM_ZEV_noControl}
\begin{split}
& \mathbf{ZEM}=\mathbf{r}_f-\mathbf{r}_{nc}\\
& \mathbf{ZEV}=\mathbf{v}_f-\mathbf{v}_{nc}
\end{split}
\end{align}
where $\textbf{r}_{nc}$ and $\textbf{v}_{nc}$ are, respectively, the position and velocity at the end of mission if \textit{no control action} is given from the considered time onward. It should be noted that using the formulation in \ref{eq:classical_ZEM_ZEV}, which will be called classical ZEM/ZEV from now on, can result in valid trajectories even for cases when the generalized acceleration term is arbitrary. In these types of environment however, using a definition of ZEM and ZEV as in \ref{eq:ZEM_ZEV_noControl}, the control gains that solve the \textit{optimal} problem are no longer the ones in \ref{eq:classical_ZEM_ZEV}. This leads to the definition of the \textit{Generalized-ZEM/ZEV} algorithm \cite{Guo:2013:ZMZV_generalized}, which is valid in any environment and will be used as starting point for the development of the proposed adaptive algorithm:
\begin{equation}
\label{eq:generalized_ZEM_ZEV}
\mathbf{a}=\dfrac{K_R}{t_{go}^2}\mathbf{ZEM} + \dfrac{K_V}{t_{go}}\mathbf{ZEV}
\end{equation}



\subsection{Reinforcement learning}
\label{sec:machine_learning}
Reinforcement Learning (RL) can be conceived as the formalization of learning by trial and error: it is based on the idea that a machine can autonomously learn the optimal behavior, or policy, to carry out a particular task, given the environment, by maximizing (or minimizing) a cumulative reward (or cost). RL algorithms work on systems that are formalized as \textit{Markov Decision Processes} \cite{Sutton:1998:sutton_reinforcement_learning,Sutton:2000:PG_convergence,Silver:2014:deterministic_PG}.


\subsubsection{Markov decision processes}
\label{sec:MDP}
The reinforcement learning problem is generally modeled as a \textit{Markov Decision Process} (MDP) which is composed by: a state space $X$, an action space $U$, an initial state distribution with density $p_1(x_1)$ representing the initial state of the system, a transition dynamics distribution with conditional density 
\begin{equation}
p(x_{t+1}|x_t,u_t)=\int_{x_{t+1}}f(x_t,u_t,x')dx'
\end{equation}
representing the dynamic relationship between a state and the next, given action $u$ and a \textit{reward} function $r$: $S\times U \rightarrow R$ that depends in general on the previous state, the current state and the action taken.  It should be noted that if the dynamics of the system is considered completely deterministic, this probability is always $0$ except when action $u_t$ brings the state from $x_t$ to $x_{t+1}$. The reward function \textit{r} is assumed to be bounded. A \textit{policy} is used to select actions by the agent given a certain state. The policy is stochastic and denoted by $\pi_\theta$ : $X\rightarrow P(U)$ where $P(U)$ is the set of probability measures of $U$,  $\mathbf{\theta} \in \mathbb{R}^n$ is a vector of $n$ parameters and $\pi_\theta (u_t|x_t)$ is the probability of selecting action $u_t$ given state $x_t$. The agent uses the policy to interact with the MDP and generate a trajectory made of a sequence of states, actions and rewards. The return
\begin{equation}
r_t^\gamma=\sum_{k=t}^{\infty} \gamma^{k-t}r(x_k,u_k)
\end{equation}
is the discounted reward along the trajectory from time step $t$ onward, with $0<\gamma \leq 1$. The agent's goal is to obtain a policy that maximizes the discounted cumulative reward from the start state to the end state, denoted by the performance objective $J(\pi)=\mathbb{E}[r_1^\gamma|\pi]$. By denoting the density at state $x'$ after transitioning for $t$ time steps from state $x$ by $p(x\rightarrow x',t,\pi)$ and the discounted state distribution by 
\begin{equation}
\rho^\pi(x'):=\int_{X} \sum_{t=1}^{\infty}\gamma^{t-1}P_1(x)p(x\rightarrow x',t,\pi) dx
\end{equation}
The performance objective can then be written as an expectation: 
\begin{equation}
J(\pi_\theta)=\int_{X}\rho^\pi(x) \int_{U}\pi_\theta(x,u)r(x,u)dudx=
\mathbb{E}_{x\sim\rho^{\pi},u\sim\pi_\theta}\left[r(x,u)\right]
\end{equation}
where $\mathbb{E}_{x\sim\rho^{\pi}}$ denotes the expected value with respect to discounted state distribution $\rho(x)$.

During training, the agent will have to estimate the reward-to-go function $J$ for a given policy $\pi$: this procedure is called \textit{policy evaluation}. The resulting estimate of $J$ is called value function. The latter may depend either on the state or both on state and action, yielding two different possible definitions. The state value function
\begin{equation}
V^\pi(x)=\mathbb{E}\left[ \sum_{k=0}^{\infty} \gamma^k r_{k+1} \vert x_0=x,\pi  \right]
\end{equation}
only depends on the state $x$. The state-action value function
\begin{equation}
Q^\pi(x,u)=\mathbb{E}\left[ \sum_{k=0}^{\infty} \gamma^k r_{k+1} \vert x_0=x, u_0=u,\pi  \right]
\end{equation}
depends on the state $x$ but also on the action $u$. The relationship between the two is:
\begin{equation}
V^\pi(x)=\mathbb{E}\left[ Q^\pi(x,u) | u \sim \pi(x,\cdot) \right]
\end{equation}
The above mentioned $V$ and $Q$ in recursive form become:
\begin{equation}
V^\pi(x)=\mathbb{E}\left[ r(x,u,x')+\gamma V^\pi(x')  \right]
\end{equation}
and
\begin{equation}
Q^\pi(x,u)=\mathbb{E}\left[ r(x,u,x')+\gamma Q^\pi(x',u')  \right]
\end{equation}
which are called \textit{Bellman Equations}.
Optimality for both $V^\pi$ and $Q^\pi$ is governed by the \textit{Bellman optimality equation}. Let $V^*(x)$ and $Q^*(x,u)$ be the optimal value and action-value functions respectively, the corresponding Bellman optimality equations are:
\begin{align}
\begin{split}
V^*(x) &=\max_u \mathbb{E} \left[ r(x,u,x')+\gamma V^*(x')  \right]\\
Q^*(x,u) &=\mathbb{E} \left[ r(x,u,x')+\gamma \max_{u'} Q^*(x',u')  \right]
\end{split}
\end{align}
The goal of reinforcement learning is to find the policy $\pi$ that maximizes $V^\pi$, $Q^\pi$ or  $J(\pi_\theta)$, or in other words, find $V^*$ or $Q^*$ that satisfy the Bellman optimality equation.


\subsubsection{Stochastic policy gradient theorem}
\label{sec:Stoc_pol_grad}
Policy gradient algorithms are among the most popular classes of continuous action and state space reinforcement learning algorithms. The fundamental idea on which they are based on is to adjust the parameters $\theta$ of the policy $\pi_\theta$ in the direction of the performance objective gradient $\nabla_\theta J(\pi_\theta)$. The biggest challenge is to compute effectively the gradient $\nabla_\theta J(\pi_\theta)$ so that at each iteration the policy becomes better than the one at the previous iteration. It turns out, from the work by Williams \cite{Williams:1992:sutton_reinforcement_learning} who theorized the REINFORCE algorithms, that the gradient of the performance objective can be \textit{estimated} using samples from experience, so without actually computing it and without a complete knowledge of the environment (sometimes referred to as \textit{model-free} algorithms). A direct implication of \cite{Williams:1992:sutton_reinforcement_learning} is the \textit{policy gradient theorem}:
\begin{multline}
\label{eq:policyGradient}
\nabla_\theta J(\pi_\theta)=\int_{X}\rho^\pi(x) \int_{U} \nabla_\theta \pi_\theta(u|x)Q^\pi(x,u)dudx = \\
= \mathbb{E}_{x\sim\rho^{\pi},u\sim\pi_\theta}\left[  \nabla_\theta \log \pi_\theta(u|x)Q^\pi(x,u) \right]
\end{multline}
where $Q(x,u)$ is the state-action value function expressing the expected total discounted reward being in state $x$ taking action $u$. The theorem is important because it reduces the computation of the performance gradient, which could be hard to compute analytically, to an expectation that can be estimated using a sample-based approach. It is important to note that this estimate is demonstrated to be unbiased so it assures that a policy is at least as good as the one in the previous iteration. Once $\nabla_\theta J(\pi_\theta)$ is computed, the policy update is simply done in the direction of the gradient
\begin{equation}
\theta_{k+1}=\theta_k+\alpha_k \nabla_\theta J_k
\end{equation}
where $\alpha$ is the learning rate and is supposed to be bounded.

One important issue to be addressed is how to estimate the $Q$ function effectively; all of the above is in fact valid in the case $Q$ represent the true action-value function. In case of continuous action and states spaces, obtaining an unbiased estimate of this is difficult. One of the simplest approach is to use the single sample discounted return $r_t^\gamma$ to estimate $Q$ which is the idea behind the REINFORCE algorithm \cite{Williams:1992:sutton_reinforcement_learning}. This is demonstrated to be unbiased\footnote{The discounted return is unbiased because it comes directly from experience and no approximation is introduced.}, but the variance is high, which leads to slow convergence. One way of estimating the action-value function in a way that reduces the variance while keeping the error contained is the introduction of a critic in the algorithm.


\subsubsection{Stochastic actor-critic algorithm}
\label{sec:stoch_act_critic}
The \textit{actor-critic} is a widely used architecture based on policy gradient. It consists of two major components. The actor adjusts the parameters $\theta$ of the stochastic policy $\pi_\theta(x)$ by stochastic gradient ascent (or descent). The critic evaluates the goodness of the generated policy by estimating some kind of value function. If a critic is present, instead of the true action-value function $Q^\pi(x,u)$, an estimated action-value function $Q^\omega(x,u)$ is used in equation \ref{eq:policyGradient}. Provided, in fact, that the estimator is compatible with the policy parametrization, meaning that
\begin{equation}
\frac{\partial Q^w(x,u)}{\partial w}=\frac{\partial \pi(x,u)}{\partial \theta} \frac{1}{\pi (x,u)}
\end{equation}
then
$Q^w(x,u)$ can be substituted to $Q^\pi(x,u)$ in \ref{eq:policyGradient} and the gradient would still assure improvement by moving in that direction. It is important to note that $Q^w(x,u)$ is required to have zero mean for each state: $\sum_{u}\pi(x,u)Q^w(x,u)=0, \quad \forall x \in X$. In this sense it is better to think of $Q^w(x,u)$ as an approximation of the \textit{advantage function} $A^\pi(x,u)=Q^\pi(x,u)-V^\pi(x)$ rather than $Q^\pi(x,u)$. This is in fact what will be used in the following.

In general introducing an estimation on the action-value function may introduce bias but the overall variance of the method is decreased which ultimately leads to faster convergence. The critic goal is to estimate the action-value function, providing a better estimate of the expectation of the reward with respect to using the single sample reward-to-go given state $x$ and action $u$. This happens because the action-value function is estimated from an average over all the samples, not just from the ones belonging to a single trajectory. This will become clearer in section \ref{sec:adaptive_ZEM_ZEV} where the details of the algorithm will be discussed. It should be noted that both the critic and the deterministic part of the policy are represented by a Single Layer Feedforward Network (SLFN). Specifically the critic is represented by an Extreme Learning Machine which is a particular instance of them and will be presented in the following section.


\subsection{Extreme learning machines}
\label{sec:ELM}
Extreme Learning Machines (ELM) are a particular kind of Single Layer Feedforward Networks (SLFN) with a single layer of hidden neurons which do not make use of back-propagation as learning algorithm. Backpropagation is a multiple step iterative process; ELM instead uses a learning method which allows for learning in a single step. The concepts behind ELM had already been in the scientific community for years before Huang theorized and formally introduced them as Extreme Learning Machines in 2004 \cite{Huang:2011:ELM_survey,Huang:2015:ELM}. According to their creator, they can produce very good results with a learning time that is a fraction of the time needed for algorithms based on back-propagation.

\begin{figure}
	\centering
	\includegraphics[width=.8\textwidth]{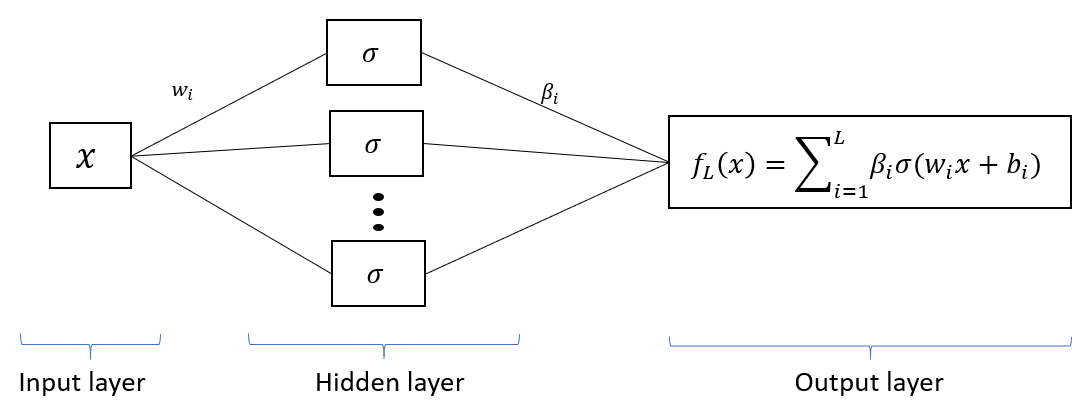}
	\caption{Single layer feedforward network}
	\label{fig:elm}
\end{figure}

Consider a simple SLFN, the universal approximation theorem states that any continuous target function $f(x)$ can be approximated by SLFNs with a set of hidden nodes and appropriate parameters. Mathematically speaking, given any small positive $\epsilon$, for SLFNs with enough number of neurons $L$, it is verified that:
\begin{equation}
\label{eq:SLFN_condition}
\Vert  f_L(x)-f(x)  \Vert < \epsilon
\end{equation}
where 
\begin{equation}
f_L(x)=\sum\limits_{i=1}^L \beta_i h_i(x)=\mathbf{H}(x) \mathbf{\beta}
\end{equation}
is the output of the SLFN, $\mathbf{\beta}$ being the output weights matrix and $\mathbf{H}(x)=\sigma(\textbf{W}x+\textbf{b})$ the output of the hidden layer for input $x$, with \textbf{W} and \textbf{b} being the input weights and biases vectors respectively, $\sigma$ is the activation function of the hidden neurons. A representation of an SLFN can be seen in Figure \ref{fig:elm}. 
In conventional SLFN, input weights $w_i$, biases $b_i$ and output weights $\beta_i$ are learned via backpropagation\footnote{Backpropagation is an optimization technique based on the concept of updating iteratively weights and biases of a neural network according to the gradient of the loss function to be minimized. It is called backpropagation because the error is calculated at the output and distributed back through the network layers. Details in \cite{Hagan:1994:backProp}}. ELM are a particular type of SLFN that have the same structure but only $\beta_i$ are learned, while input weights and biases are assigned randomly at the beginning of training without the knowledge of the training data and are never changed. It is demonstrated that, for any randomly generated set $\left\lbrace \textbf{W},\textbf{b} \right\rbrace$ of input weights and biases,
\begin{equation}
\lim_{L\rightarrow\infty} \Vert  f(x)-f_L(x)  \Vert = 0
\end{equation}
holds if the output weights matrix $\mathbf{\beta}$ is chosen so that it minimizes \ref{eq:SLFN_condition}, which is equivalent to saying that it minimizes the loss function $\Vert  f(x)-f_L(x)  \Vert$. Equation \ref{eq:SLFN_condition} after some manipulation, becomes
\begin{equation}
\label{eq:ELM_loss}
\Vert  \mathbf{H}\beta - \mathbf{Y} \Vert  < \epsilon
\end{equation}
where $\mathbf{Y}=[\mathbf{y_1},...,\mathbf{y_N}]^T$ are the target labels and $\mathbf{H}=[\mathbf{h^T(x_1)},...,\mathbf{h^T(x_N)}]$ the hidden layer output. Given $N$ training samples $ \lbrace x_i,y_i \rbrace_{i=1}^N$, the training problem is reduced to:
\begin{equation}
\label{eq:leastSquare}
\mathbf{H} \mathbf{\beta}=\mathbf{Y}
\end{equation}
The output weights are then simply:
\begin{equation}
\mathbf{\beta}=\mathbf{\tilde{H}}\mathbf{Y}
\end{equation}
Where $\mathbf{\tilde{H}}$ is the Moore-Penrose generalized inverse matrix\footnote{Used here because the numbers of neurons and samples are not equal so the system is not squared.} of $\mathbf{H}$. This is demonstrated to minimize the loss \ref{eq:ELM_loss} given a large enough sets of training points and neurons. This is another way of saying that $\mathbf{\tilde{H}}$ are the weights that represent the minimum norm least square solution of \ref{eq:leastSquare}. This will be used in the actor-critic algorithm and will be explained in section \ref{sec:adaptive_ZEM_ZEV}.


\section{Adaptive-ZEM/ZEV algorithm}
\label{sec:adaptive_ZEM_ZEV}

\begin{figure}
	\centering
	\includegraphics[width=.55\textwidth]{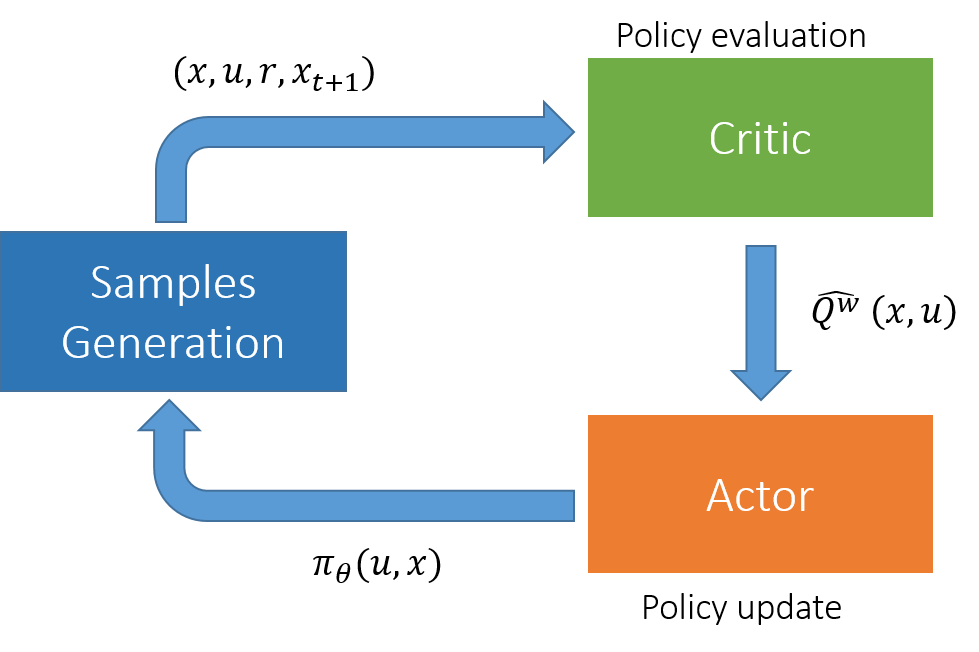}
	\caption{Schematic representation of the actor-critic algorithm}
	\label{fig:A_ZEM_ZEV}
\end{figure}

The Adaptive-ZEM/ZEV (A-ZEM/ZEV) is based on the idea of learning the parameters $K_R$, $K_V$ and the time of flight $T_f$, which is related to the time-to-go $t_{go}$ of the generalized ZEM/ZEV algorithm in Equation \ref{eq:generalized_ZEM_ZEV}. The overall idea is that the guidance gains and the time-to-go can be adapted during the powered descent phase to satisfy specific constraints while maintaining quasi-fuel optimality and close-loop characteristics. The guidance adaptation is achieved by using a customization of the actor-critic algorithm described in Section \ref{sec:machine_learning}.  More specifically, we have developed a fast RL framework  based on a combination of the REINFORCE algorithm \cite{Williams:1992:sutton_reinforcement_learning} and a critic network based on Extreme Learning Machines for estimating the value function. The goal is to show that learning of the adaptive generalized ZEM/ZEV guidance can occur fastly and effciently.
The proposed RL-based learning algorithm can be broken down in three major blocks:
\begin{enumerate}
	\item Samples generation
	\item Critic neural network fitting
	\item Policy update
\end{enumerate}
A high level schematic representation of the algorithm can be seen in Figure \ref{fig:A_ZEM_ZEV}.
Overall, at each global iteration, a batch of sample trajectories is generated giving a set of states, actions, costs and next states $(x,u,c,x_{t+1})$. These are then fed to the critic that outputs an approximation of the expected cost-to-go given a particular action and state $\hat{Q}^w(x,u)$ that is then used to update the policy by the actor. Importantly, we do not aim at learning the full guidance policy, which is represented by the generalized ZEM/ZEV algorithm (Eq. \ref{eq:generalized_ZEM_ZEV}). Here, we call \textit{policy} specifically $K_R$, $K_V$ and $T_f$ as function of the lander state and parametrized/approximated by a neural networks (see Figure \ref{fig:policy} for a description of the policy network). Indeed, the parameters of such networks are learned during the training phase. Details of the three phases are reported in the next sections.

\subsection{Samples generation}
\label{sec:sample_gen}
At each global iteration, a batch of trajectories are generated by letting the agent interact with the environment using policy $\pi_\theta(u|x)$, which is a representation of the guidance gains in equation \ref{eq:generalized_ZEM_ZEV}, giving a series of samples $(x_{i,t},u_{i,t},c_{i,t},x_{i,t+1})$, where $i$ represents the trajectory number and $t$ is the time-step along that trajectory. At the start of each episode, the time of flight $T_f$ is sampled using the policy and kept constant for the entire episode. The starting position is randomly chosen by sampling a gaussian distribution around the nominal starting position. This ensures exploration of the state space around the nominal starting state and also avoids singularities in the policy evaluation step\footnote{The critic network works well only if each state is associated with a single value. If each episode starts from the exacts same position, there are equal states associated with different costs, which makes the regression perform poorly.}. The time is discretized in a fixed number of time steps: at the beginning of each time step the policy is sampled and $K_R$ and $K_V$ obtained, the acceleration command calculated with \ref{eq:generalized_ZEM_ZEV}, and the equations of motion integrated forward in time. The acceleration command is kept constant during the time interval. The cost, whose value depend on the particular case addressed, is assigned at each time step. It should be noted that here the \textit{reward} in the definitions in Section \ref{sec:machine_learning} in substituted with a \textit{cost} for reasons that will become clearer later. Importantly, the whole machinery described in Section \ref{sec:machine_learning} is valid also in case a cost is used to evaluate actions instead of a reward. The final time for each episode is also fixed and the agent runs until the end time is reached unless an impact with the ground is detected in which case the episode ends.


\subsubsection{Policy}
\label{sec:policy}
\begin{figure}
	\centering
	\includegraphics[width=.55\textwidth]{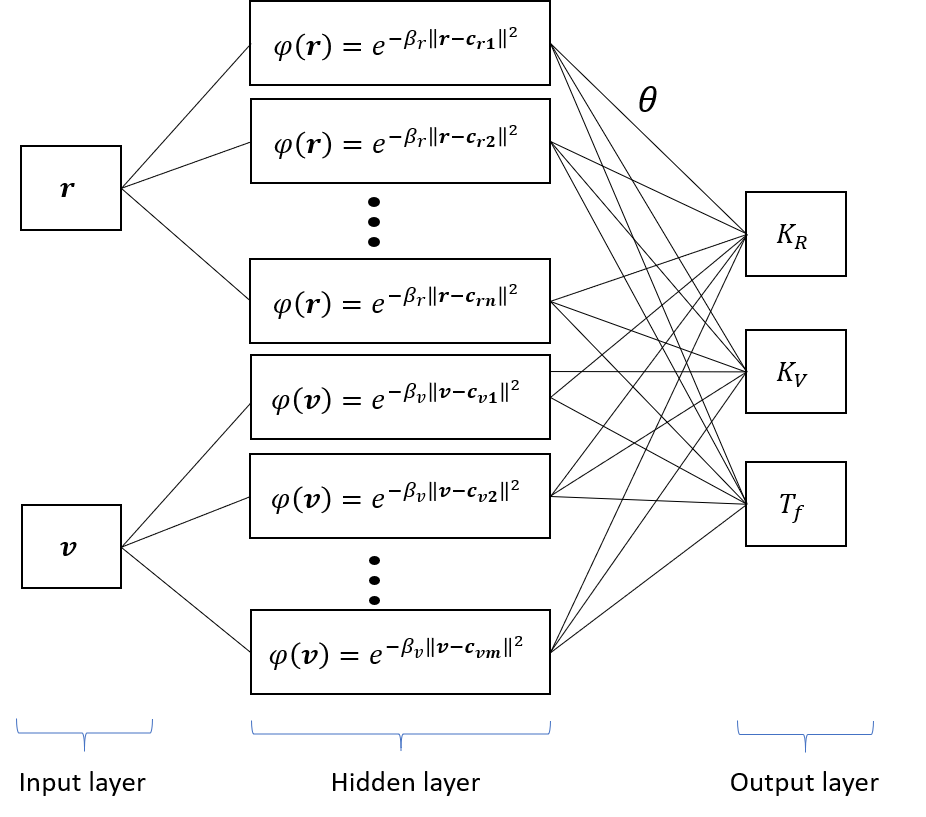}
	\caption{Policy neural network}
	\label{fig:policy}
\end{figure}

The policy is described by a gaussian distribution with fixed variance $\sigma^2$ and variable mean from which actions are sampled. The mean is parametrized over a certain weight vector $\theta$ which is learned through gradient descent. The stochasticity  of the policy is essential for learning because 1) it enables exploration of the action space and 2) the machinery developed for stochastic policy gradient can then be applied. Since the parameters of the guidance algorithm to learn are three, i.e. $K_R$, $K_V$ and $T_f$, the policy is subdivided in three separate parts and parametrized with ($\theta_{K_R}$, $\theta_{K_V}$ and $\theta_{T_f}$). The policy can be formally expressed as:
\begin{align}
\label{eq:policy}
K_R &= \pi_{\theta_{K_R}}=\mathcal{N}(\mu_{K_R},\sigma^2)\\
K_V &= \pi_{\theta_{K_V}}=\mathcal{N}(\mu_{K_V},\sigma^2)\\
T_f &= \pi_{\theta_{T_f}}=\mathcal{N}(\mu_{T_f},\sigma^2)
\end{align}
where:
\begin{align}
\mu_{K_R} &= \mathbf{ \phi(x)}^T \theta_{K_R}\\
\mu_{K_V} &= \mathbf{ \phi(x)}^T \theta_{K_V}\\
\mu_{T_f} &= \mathbf{ \phi(x)}^T \theta_{T_f}
\end{align}
$\mathbf{ \phi(x)}$ is the vector of feature functions evaluated in state \textbf{x} and $\theta_{K_R}$, $\theta_{K_V}$ and $\theta_{T_f}$ are the weight vectors associated with each output. Note that the mean values learned during the training phase are employed in the generalized ZEM/ZEV algorithm. The features are comprising two sets of three dimensional radial basis functions (RBF) with centers distributed evenly across the position and velocity spaces. They are represented by the expression:
\begin{align}
& \phi(\mathbf{r})=e^{-\beta_R \Vert\mathbf{r}-\mathbf{c_r}\Vert^2}\\
& \phi(\mathbf{v})=e^{-\beta_V \Vert\mathbf{v}-\mathbf{c_v}\Vert^2}
\end{align}
with $\beta_R$ and $\beta_V$ being constant parameters related to the variance of the radial functions which is set accordingly to the particular case, $\mathbf{r}$ and  $\mathbf{v}$ being respectively the position and velocity and $\mathbf{c_r}$ and  $\mathbf{c_v}$ the centers of the RBFs. The centers are generated by dividing the state space of the problem in a set of intervals, thus creating a grid of equally spaced points in the position and velocity spaces. The deterministic part of this policy can be seen as a neural network with two three-dimensional inputs ($\mathbf{r}$,$\mathbf{v}$), a single hidden layer of neurons with radial basis activation functions and a three dimensional output layer ($K_R$, $K_V$, $T_f$). A scheme can be seen in Figure \ref{fig:policy}. The parameters $\theta$ are the weights that multiplied by the features give the output, which is the mean of the stochastic policy as stated above.


\subsection{Critic neural network}
\label{sec:critic}
One key part of the algorithm is the fitting of the neural network that approximates the value function. As explained in \ref{sec:stoch_act_critic}, in actor-critic algorithms the expectation in equation \ref{eq:policyGradient} is not computed exactly, but it is rather expressed using an approximated value function $Q^w(x,u)$. Here, we employ the advantage function $A^\pi(x,u)=Q^\pi(x,u)-V^\pi(x)$ rather than $Q^\pi(x,u)$. The approximated advantage function can be rewritten, using the definition of $Q$, as function of $V$ only:
\begin{equation}
\label{eq: value_equation}
Q^w(x,u) = \hat{A}^{\pi}(u,x)= \hat{Q}^\pi(x,u)-\hat{V}^\pi(x)= \\
r(x,u)+\hat{V}^\pi(x_{t+1})-\hat{V}^\pi(x)
\end{equation}
where $\hat{A}^{\pi}(u,x)$, $\hat{Q}^{\pi}(u,x)$ and $\hat{V}^{\pi}(x)$ are the approximated versions of $A^{\pi}(u,x)$, $Q^{\pi}(u,x)$ and $V^{\pi}(x)$.  Clearly, in order to compute the approximated advantage function, only $\hat{V}^{\pi}(x)$ must be obtained. The latter is done by modeling the value function via a single layer forward networks with the following sigmoid activation function
\begin{equation}
\sigma(s_i)=\frac{1}{1+e^{-s_i}} \quad \text{with} \quad s_i=w_i x + b_i
\end{equation}

The SLFN is used as a function approximator that maps the inputs, in this case the 6D states, into the scalar representing the discounted cost and trained at each step using ELM theories as introduced earlier. The latter is done by generating at each global iteration step, a training set on which the SLFN is trained using the training algorithm described in Section \ref{sec:ELM}. There are normally two ways to define this training set referring to two different types of methods:
\begin{itemize}
	\item Monte Carlo (MC): the value function is approximated at any given state $x_{i,t}$ by the return, which is the discounted cost-to-go  $y=\sum_{t'=t}^{T} \gamma^{t'-t} c(x_{i,t'},u_{i,t'})$. In this case the training set is defined by the couples:
	\begin{equation}
	\label{eq:trainingSet_MC}
	\bigg \{ \left( x_{i,t} , \sum_{t'=t}^{T} \gamma^{t'-t} c(x_{i,t'},u_{i,t'}) \right) \bigg \}
	\end{equation}
	This is an unbiased way of expressing the value function but could suffer from high variance.
	\item Temporal Difference (TD): the value function is approximated by a bootstrapped estimate of the cost-to-go, meaning that the previously fitted value function is used as an estimation of the cost-to-go from time step $t+1$ onward. The training set in this case is given by the couples:
	\begin{equation}
	\label{eq:trainingSet_TD}
	\bigg \{ \left( x_{i,t} , c(x_{i,t},u_{i,t}) + \hat{V}^\pi(x_{t+1}) \right) \bigg \}
	\end{equation}
	this way of expressing the value function introduces a bias, because the estimation of $V$ is not perfect, but reduces the variance.
\end{itemize} 
In this case the possibility of using the TD errors for value function approximation as in \ref{eq:trainingSet_TD} was explored but discarded in favor of the MC version \ref{eq:trainingSet_MC}. Here, \ref{eq:trainingSet_TD} works well only when the bias introduced by the approximation is small. In this case, in two consecutive global iterations the visited states could be very different. Consequently, the neural net approximating the value function and trained on a particular portion of the state space could lead to very big extrapolation errors. For this reason, we decided to employ the MC version of the algorithm to keep the bias contained and find other means of reducing the variance. It should be noted right away that, even if the variance is higher with respect to the bootstrapped version, it is still lower than that of the vanilla REINFORCE algorithm. Indeed, the samples come from all the generated episodes, and therefore the learned value function is an \textit{average} of the expected cost-to-go, which is a better estimate of the value function with respect to the simple sample estimate. Note also that the approximated value function is the discounted cost and not the discounted reward. This is a choice made for this particular case in which the goodness of an action is more clearly represented by a cost instead of a reward.


\subsection{Policy update}
\label{sec:actor}
During the training phase, the policy is optimized using gradient descent instead of gradient ascent. The latter requires gradient estimation to execute the policy update step. Once the value function is approximated by the critic net, it is used to estimate the gradient of the objective function $J(\pi_\theta)$. In stochastic policy gradient, the expectation in equation \ref{eq:policyGradient} is not computed directly but is  approximated by averaging the gradient over the samples. In this case a batch of trajectories is used to estimate the gradient. The expression of the approximated gradient becomes:
\begin{equation}
\nabla_\theta J(\pi_\theta) \approx \dfrac{1}{N} \sum_{i=1}^{N} \sum_{t=1}^{T} \nabla_\theta \log \pi_\theta(u_{i,t}|x_{i,t}) \hat{A}^{\pi}(u_{i,t},x_{i,t})
\end{equation}
where $N$ is the number of sample trajectories in the batch, $T$ is the number of time instants in each trajectory, $\nabla_\theta \log \pi_\theta(u|x)$ is the gradient of the log-probability of the stochastic policy which, for a gaussian policy like \ref{eq:policy}, is obtained analytically as:
\begin{equation}
\nabla_\theta \log \pi_\theta = \dfrac{\pi_\theta-\mu}{\sigma^2} \mathbf{\mathbf{\phi}(s)}
\end{equation}
Here, $\hat{A}^{\pi}(u_t,x_t)$ is the approximated advantage function described in \ref{sec:critic} and indicates how much better the action $u_{i,t}$ performs with respect to the average action. Using the advantage function generally reduces the variance (\ref{sec:stoch_act_critic}) but it relies on an approximation that introduces bias into the process. A way to reduce the effect of bias is to use the advantage function formulated as:
\begin{equation}
    \hat{A}^{\pi}_n (u_{i,t},x_{i,t}) = \sum_{t'=t}^{T} \gamma^{t'-t} c(x_{i,t'},u_{i,t'}) - \hat{V}^{\pi}(x_{i,t})
\end{equation}
which is often referred to as the Monte-Carlo formulation of the advantage function, with the discount factor being introduced as $0<\gamma<1$. This is unbiased because the real cost to go is used to estimate the action-value function but is low in variance because the average value associated to state $x_{i,t}$ is subtracted.

\begin{figure}
	\centering
	\includegraphics[width=.8\textwidth]{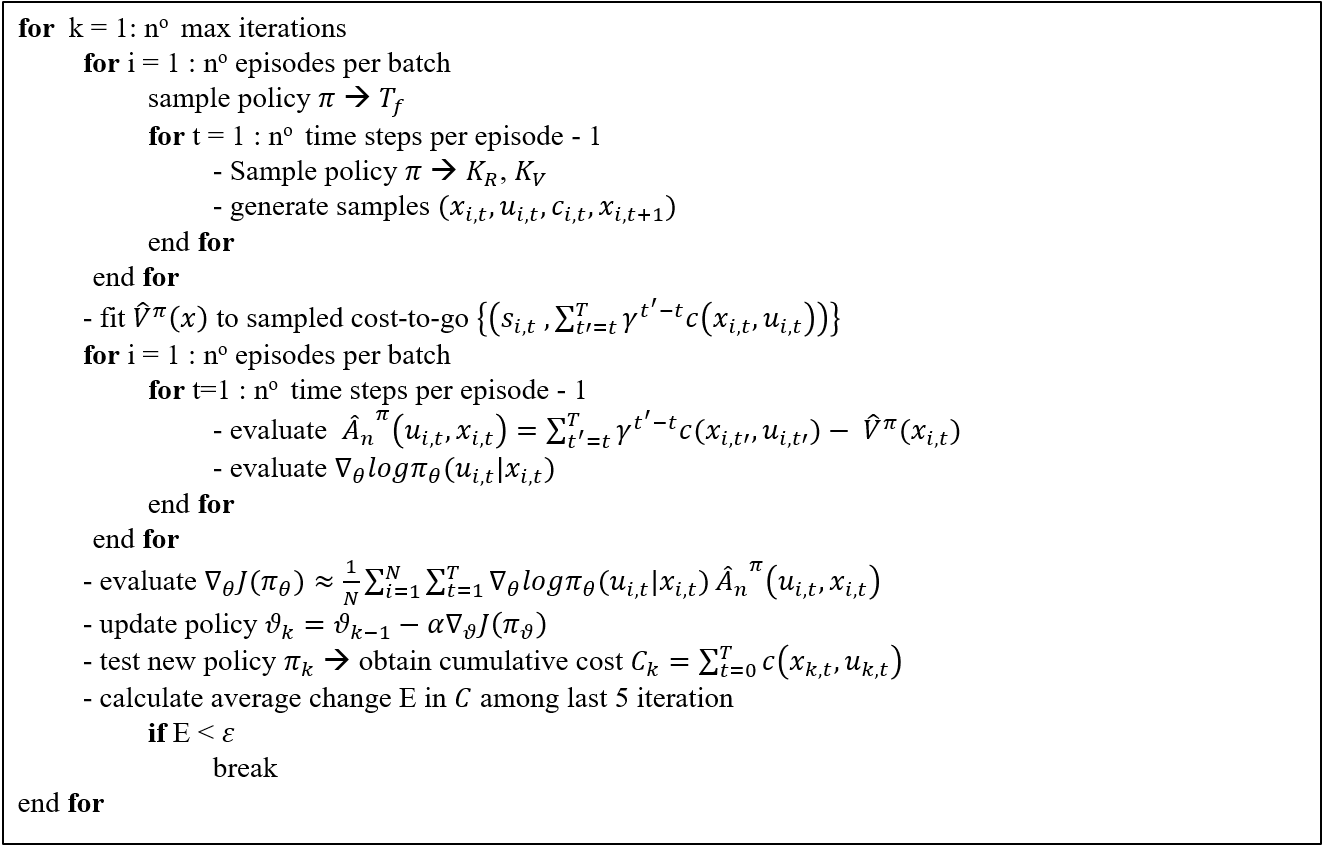}
	\caption{Summary of the A-ZEM/ZEV algorithm}
	\label{fig:A_ZEM_ZEV_codice}
\end{figure}

To implement the gradient descent algorithm, the policy parameters update is simply done by taking a step in the opposite direction of the gradient $\nabla_\theta J(\pi_\theta)$:
\begin{equation}
\theta_{k+1}=\theta_k - \alpha \nabla_\theta J(\pi_\theta)
\end{equation}
where $\alpha$ is the bounded learning rate. After each update, the algorithm is tested and the cumulative cost is computed
\begin{equation}
C_k=\sum_{t=0}^{T}c(x_t,u_t)
\end{equation}
where $k$ stands for \textit{k}-th iteration. The algorithm stops if the average cumulative cost difference among the last 5 iteration is less than a tolerance $\epsilon$ or it has reached the maximum number of iterations. A summary of the algorithm in form of pseudo-code is given in Figure \ref{fig:A_ZEM_ZEV_codice}.


\section{Numerical Results}
\label{sec:results}

To evaluate the performance of the proposed algorithm, two powered descent examples for Mars pinpoint landing are presented. The spacecraft parameters for the selected problems are the following:

\begin{equation}
    \begin{split}
        \mathbf{g} = [-3.7114 \, , \,  0 \, , \, 0]^T m/s^2 \qquad m_{dry} = 1505 \, kg \\
        m_{wet} = 1905\, kg \qquad I_{sp} = 225 \, s \qquad \bar{T} = 3.1 \, kN \\
        T_{min} = 0.3\bar{T} \qquad T_{min} = 0.8\bar{T} \qquad n = 6
    \end{split}
\end{equation}

Here, $m_{wet}$ and $m_{dry}$ are the spacecraft mass both wet and dry, respectively, and $\mathbf{g}$ is the Martian gravity vector, assumed to be constant. Additionally, $n$ is the number of thrusters, each with a full throttle capability of $\bar{T}$, limited to a thrust level between 0.3 and 0.8 at all times. The thrusters are mounted on the spacecraft body with a cant angle $\phi$ with respect to the net thrust direction $\hat{T}$. If $\bar{\mathbf{T}}$ as the magnitude of instantaneous thrust for each individual thruster, the instantaneous net thrust is:
\begin{equation}
    \mathbf{T}_n=n\bar{T}cos(\phi)\hat{\mathbf{T}}
\end{equation}

The A-ZEM/ZEV algorithm was tested on two cases: a 2D case where the spacecraft is constrained to move on the $x$-$z$ plane and a full 3D case.  The guidance gains and the final time of the adaptive algorithms are learned to ensure safe landing at a selected location of the Martian surface with minimum fuel. it is assumed that the target point is at the origin of the reference frame fixed with the Martian surface which needs to be achieved with zero velocity. In both cases a glide constraint is introduced:
\begin{equation}
    \label{eq:glide_slope}
    \theta(t) = \arctan (\frac{\sqrt{r_x(t)^2 + r_y(t)^2}}{r_z(t)}) \leq \theta_{lim} 
\end{equation}

with an angle $\theta_{lim}= 4 ^{\circ}$ with respect to the horizon. During the descent, the gains adaptation must ensure that this constraint is always satisfied. This is achieved by terminating the episodes whenever the agent violates the constraint, which also leads to an increase in cost. The cost function $c(t)$ that enforces that constraint while searching for fuel optimal solutions is defined as follows:
\begin{multline}
\label{eq:cost_function}
C(t)=w_m dm_t
+\delta(t-T_f) \left[ w_r^f \Vert r_t-r_f \Vert^2 + w_v^f \Vert v_t-v_f \Vert^2 + b_f \right]  \\
+\delta(t-t_i) \left[ w_r^i \Vert r_t-r_f \Vert^2 + b_i \right]
\end{multline}

Where $w_m$, $w_r^f$, $w_v^f$ and $w_r^i$ are weights associated with the burned mass, the end position and velocity errors and the impact point position error respectively, $t_i$ and $T_f$ are the time of impact and the final time respectively, and $b_f$ and $b_i$ are biases added at the end of episodes with $b_i>b_f>0$.  

Importantly, $b_i>b_f$ ensures that the collision-less solution has a lower cost than a solution that impacts on the constraint. Conversely, $b_f>0$ ensures that the value function close to the target does not get too close to 0. The latter may cause problems during training phase because the error introduced by the function approximator might be high relative to the actual value. It is important to note here that the introduction of the positive bias $b_i$ is what ensures that the agent is incentivised to look for a collisionless solution, enforcing the constraint in equation \ref{eq:glide_slope}.

Setting up the cost is the hardest hustle because the agent can easily fall into a local minimum due to one of the multiple terms in the cost function prevailing over the others. Since the guidance adaptation has to minimize fuel cost without violating the constraints, a careful tuning of the weights values is mandatory. Here, we have decided to add a high bias cost every time an episode ended with an impact. The latter ensures that the minimum cost is always achieved with a collision-less solution. In this fashion, we have observed that the algorithm always tries first to avoid the constraint, then to lower the fuel consumption. The introduction of this constant biases leads to discontinuous jumps in the cost profile as training progresses. The reason is that a trajectory without collisions has a much lower cost than one that collides with the constraint as $b_i>b_f$. 

For both 2D and 3D guidance simulations, the starting position for each episode was sampled from a uniform distribution around the nominal starting state. Table \ref{tab:initials} shows the initial state distributions for both cases.
\begin{table}
	\caption{Initial state distributions}
	\label{tab:initials}
	\centering
	\begin{tabular}{c | c | c | c | c | c | c} 
		\toprule
		 & $x$ (m) &  $y$ (m) & $z$ (m) & $\dot{x}$ (m/s) & $\dot{y}$ (m/s) &  $\dot{z}$ (m/s)\\
		\midrule
		   Nominal 2D & 1500 & 0 & 1500 & 100 & 0 & -60 \\
		   Nominal 3D & -500 & -1000 & 1500 & 100 & -60 & -60 \\
		   Bounds & $\pm$ 500 &	 $\pm$ 500  & $\pm$ 0 & $\pm$ 5  &  $\pm$ 5 &  $\pm$ 5 \\
		\bottomrule
	\end{tabular}
\end{table}
Table \ref{tab:hyperparameters} instead shows the values of the hyperparameters used in the definition of the cost function in \ref{eq:cost_function}.
\begin{table}
	\caption{Hyperparameters}
	\label{tab:hyperparameters}
	\centering
	\begin{tabular}{c | c | c | c | c | c} 
		\toprule
		 $w_m$ & $w_r^f$ & $w_v^f$ & $w_r^i$ & $b_i$ & $b_f$\\
		\midrule
		 0.5  & 1e-1 & 1e-1 & 5e-4 & 100 & 10 \\
		\bottomrule
	\end{tabular}
\end{table}

\begin{figure}
	\centering
	\subfloat[][Trajectory\label{fig:2d_traj1}]
	{\includegraphics[width=.99\textwidth]{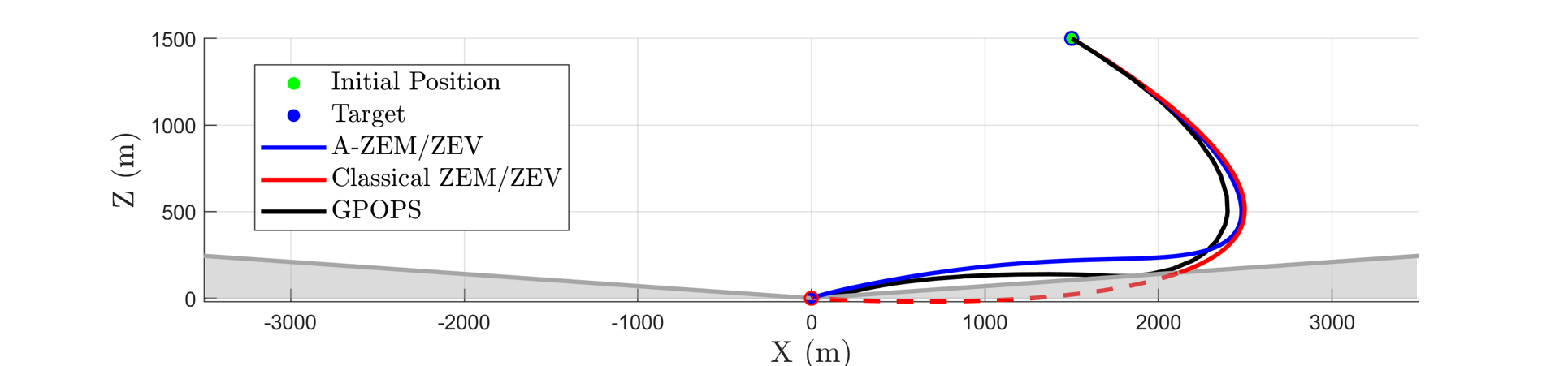}} \\
	\subfloat[][Position\label{fig:2d_pos}]
	{\includegraphics[width=.45\textwidth]{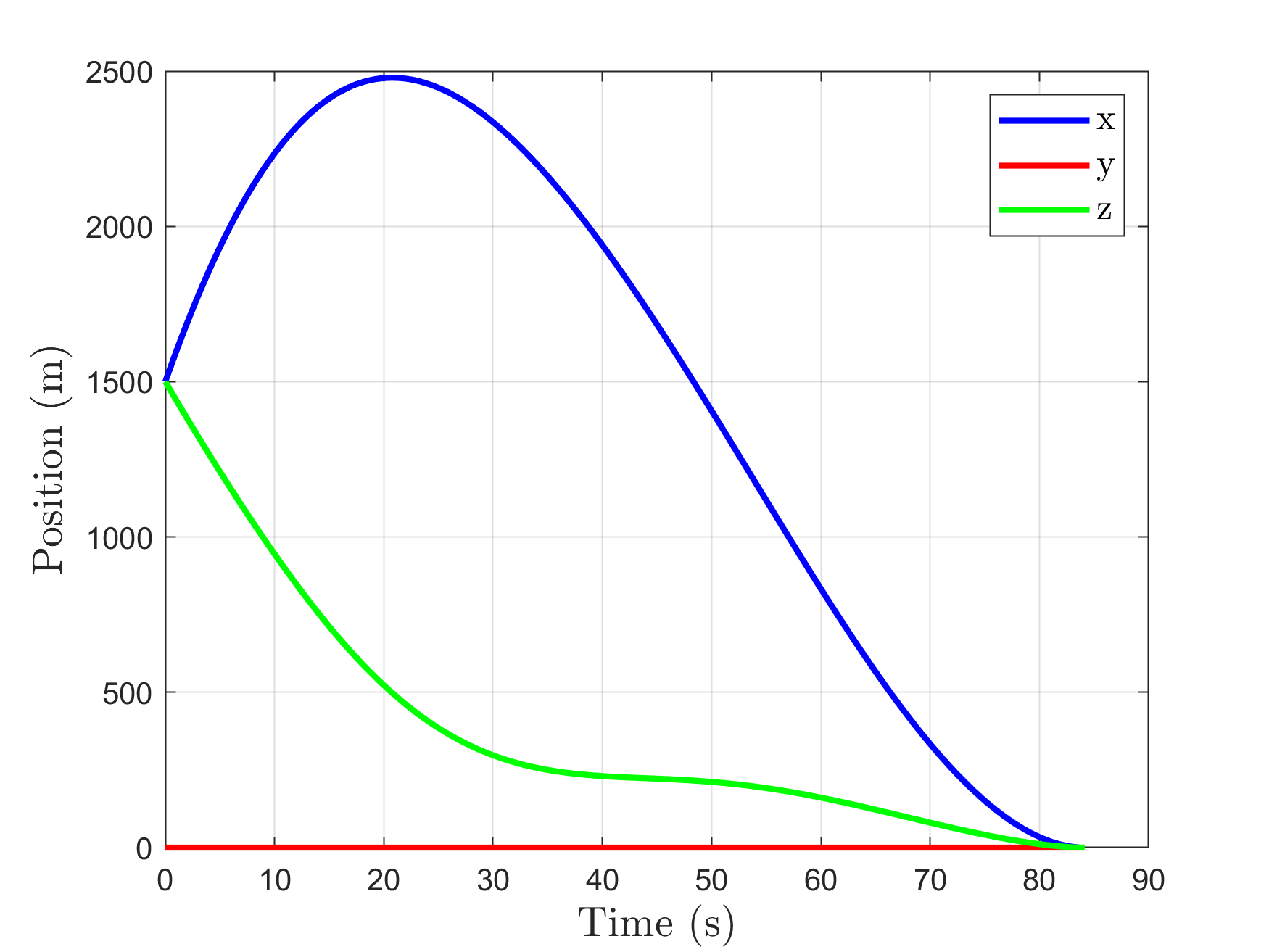}}
	\subfloat[][Velocity\label{fig:2d_vel}]
	{\includegraphics[width=.45\textwidth]{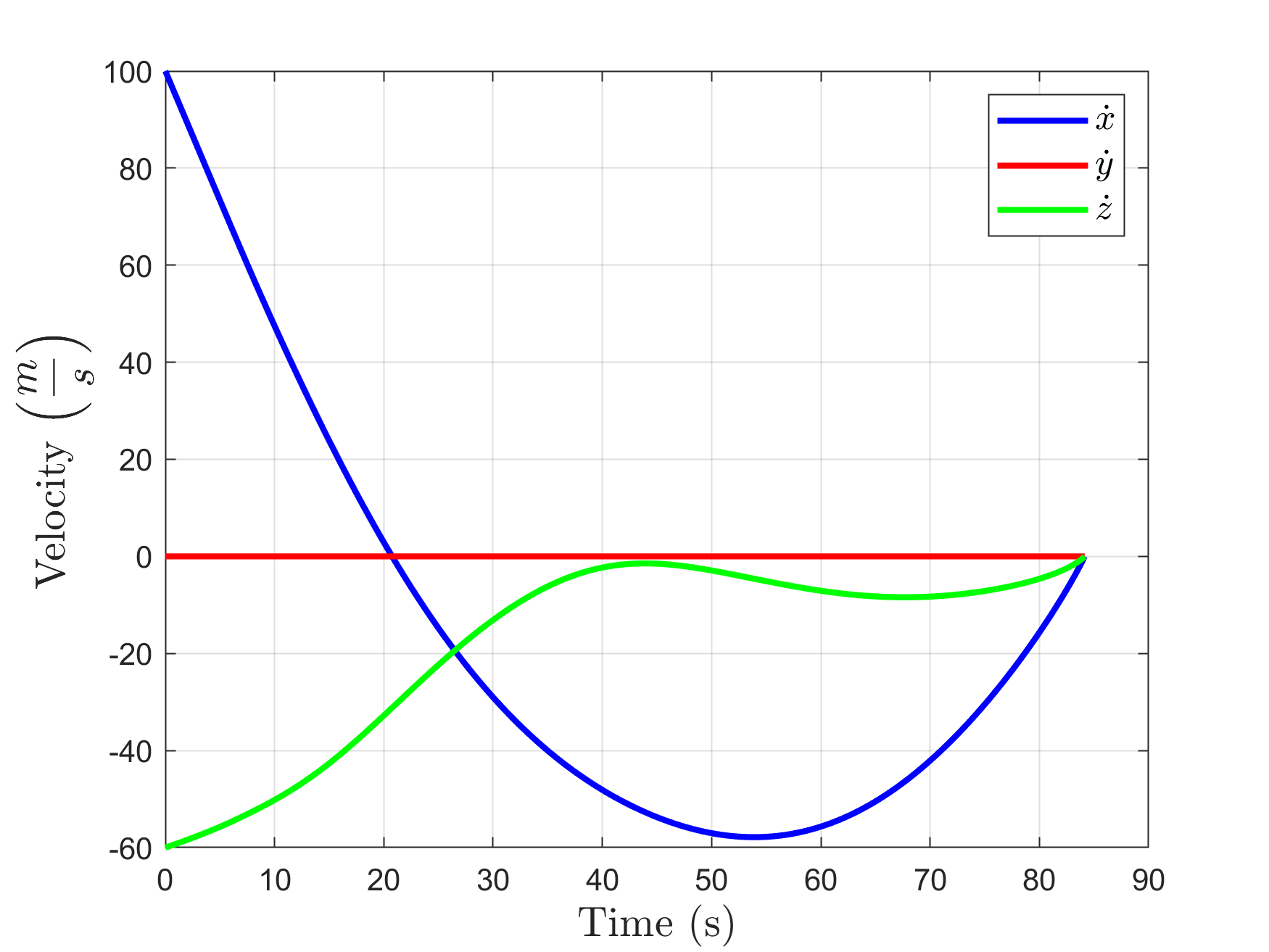}}\\
	\subfloat[][Norm of the position error with respect to the GPOPS solution.\label{fig:errors}]
	{\includegraphics[width=1\textwidth]{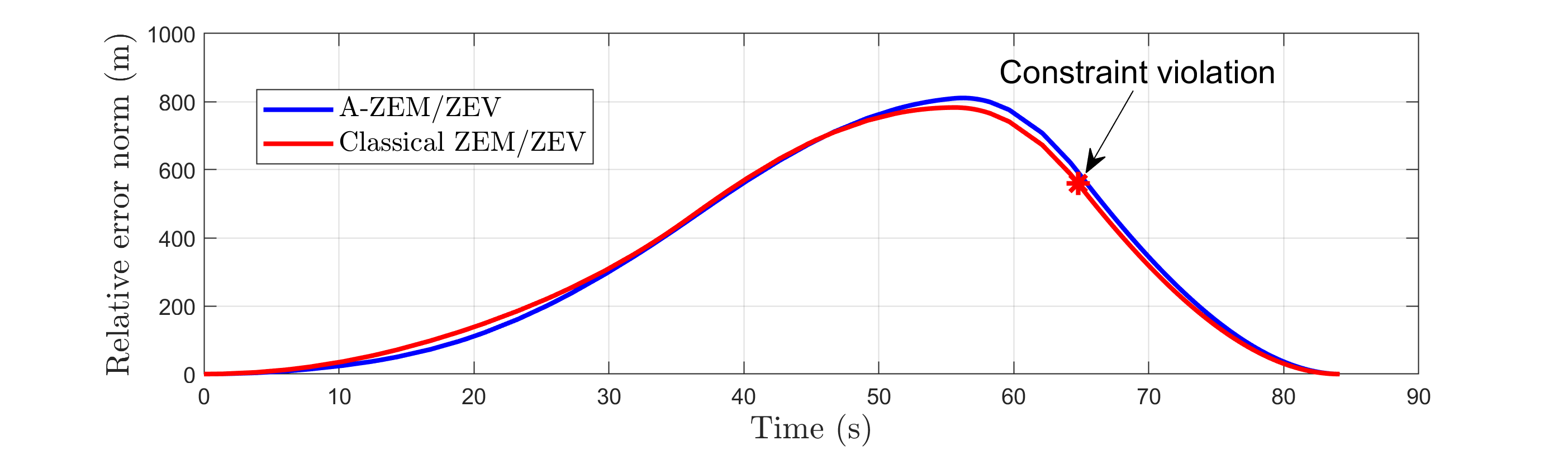}}
	\caption{$r_0=[1500 ,0, 1500]^T m$, $\dot{r_0}=[100 ,0, -60]^T m/s$. Note that although achieving the target state, the classical ZEM/ZEV violate the slope constraints. Conversely, as seen in (a) the Adaptive ZEM/ZEV guidance law does not violate the slope constraints.}
	\label{fig:2d_traj}
\end{figure}

Additionally, in the 3D case, at each iteration 25 test trajectories were generated after the policy update step. To  evaluate the performances of the current version of the policy, the initial state is selected from the same distribution used for the training samples. The latter slows down the learning process but allows to learn a policy that works for any starting state within the above mentioned distribution. In both 2D and 3D guided descents, the solutions obtained with A-ZEM/ZEV is compared with the classical ZEM/ZEV solution and a fuel optimal solution obtained with GPOPS \cite{Patterson:2014:GPOPS} (\textit{\textbf{G}eneral \textbf{P}seudospectral \textbf{OP}timal Control \textbf{S}oftware}).
The 2D guided descent scenario results are reported in Figure \ref{fig:2d_traj}. Clearly, the algorithm manages to find a solution that comply with the glide slope constraint whereas the classical-ZEM/ZEV solution violates it. Here, the ELM-based critic is trained using 80\% of the training set defined in \ref{eq:trainingSet_MC} and the rest is used for testing. Note that this step is repeated at each time-step with a different training set. The number of neurons of the ELM is set as 1/10 of the number of samples. The latter is demonstrated to work well in all situations, the most challenging ones being iterations where only some of the sample trajectories would hit the constraints and the other would arrive at the target. This condition created a stiff value function with neighbouring states having very different values given by a different future development (one hitting the constraint, the other reaching the target). Overall, A-ZEM/ZEV has shown good performance in terms of constraint avoidance. Fuel consumption is considered in the cost but there is still space for improvement as the GPOPS fuel optimal solution is not reached.

Some details on the training process are shown in Table \ref{tab:ELM}. One can appreciate that the ELM has a very short learning time if compared to the overall iteration time. This ensures that most of the computational time is spent generating the sample trajectories and updating the policy.

\begin{table}[htbp]
	\fontsize{10}{10}\selectfont
    \caption{RL performance}
   \label{tab:ELM}
        \centering 
   \begin{tabular}{c  | p{1cm} | p{2cm} | p{2cm} | p{2cm} | p{2cm} } 
      \hline 
      {\bfseries Case}   & {\bfseries $N^o$ iterations}  & {\bfseries Total training time (hours)}  & {\bfseries Average iteration time (s)}  & {\bfseries Average critic training time (s)}  & {\bfseries  Average critic NRMSE\footnote{Normalized Root Mean Squared Error}}\\
      \hline 
      2D     & 503  & 1.78 & 12.72 & 7.849e-2 & 2.860e-1 \\
      3D     & 804  & 20.68 & 92.58 & 2.727e-1 & 1.742e-1 \\
      \hline
   \end{tabular}
\end{table}


\begin{figure}
	\centering
	\subfloat[][Thrust\label{fig:2d_thrust}]
	{\includegraphics[width=.45\textwidth]{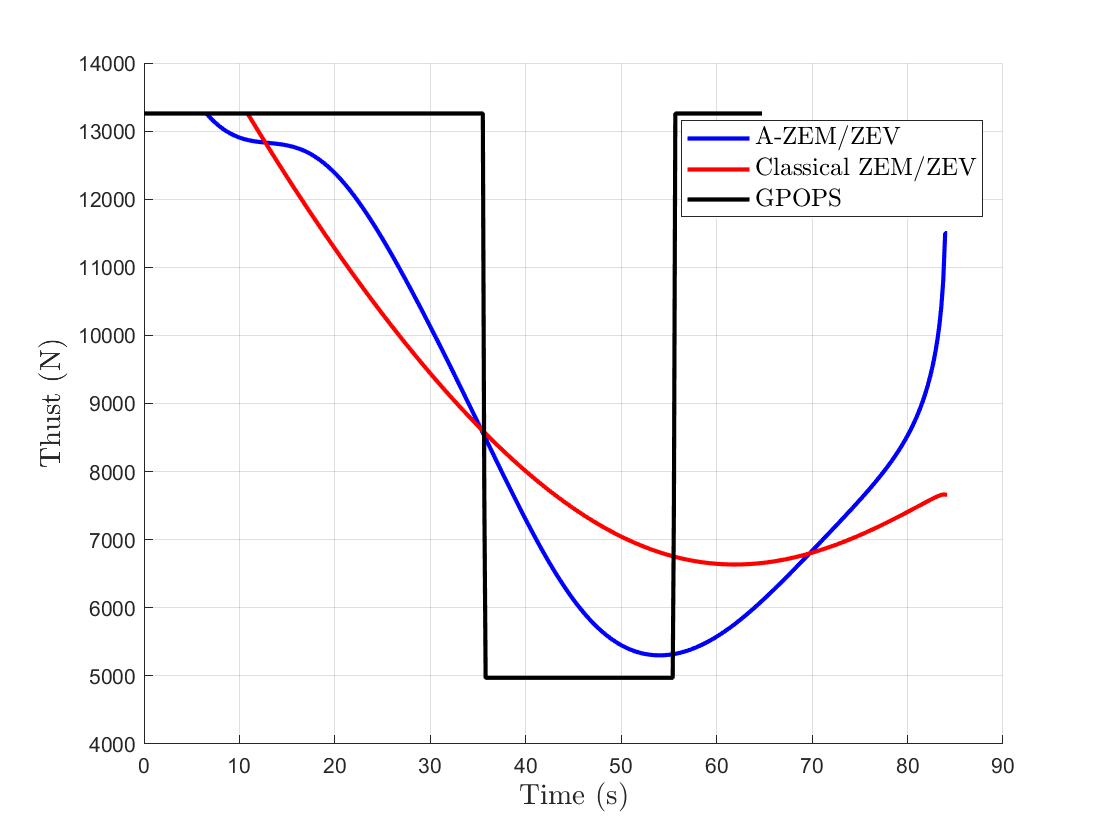}}
	\subfloat[][Spacecraft mass\label{fig:2d_mass}]
	{\includegraphics[width=.45\textwidth]{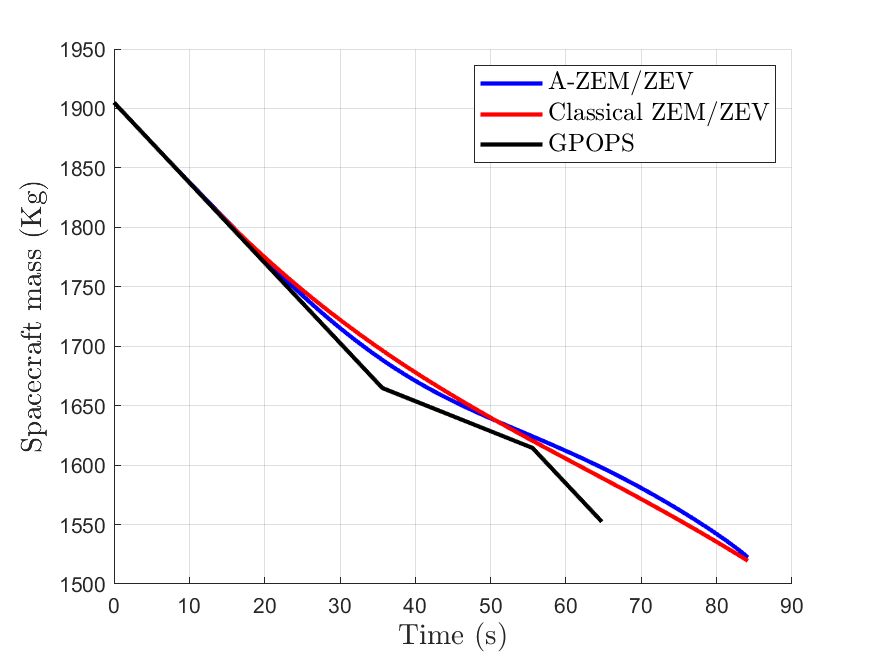}}
	\caption{Thrust, mass and guidance gains for 2D case}
	\label{fig:2d_mass_thrust_gains}
\end{figure}
\begin{figure}
	\centering
	{\includegraphics[width=.60\textwidth]{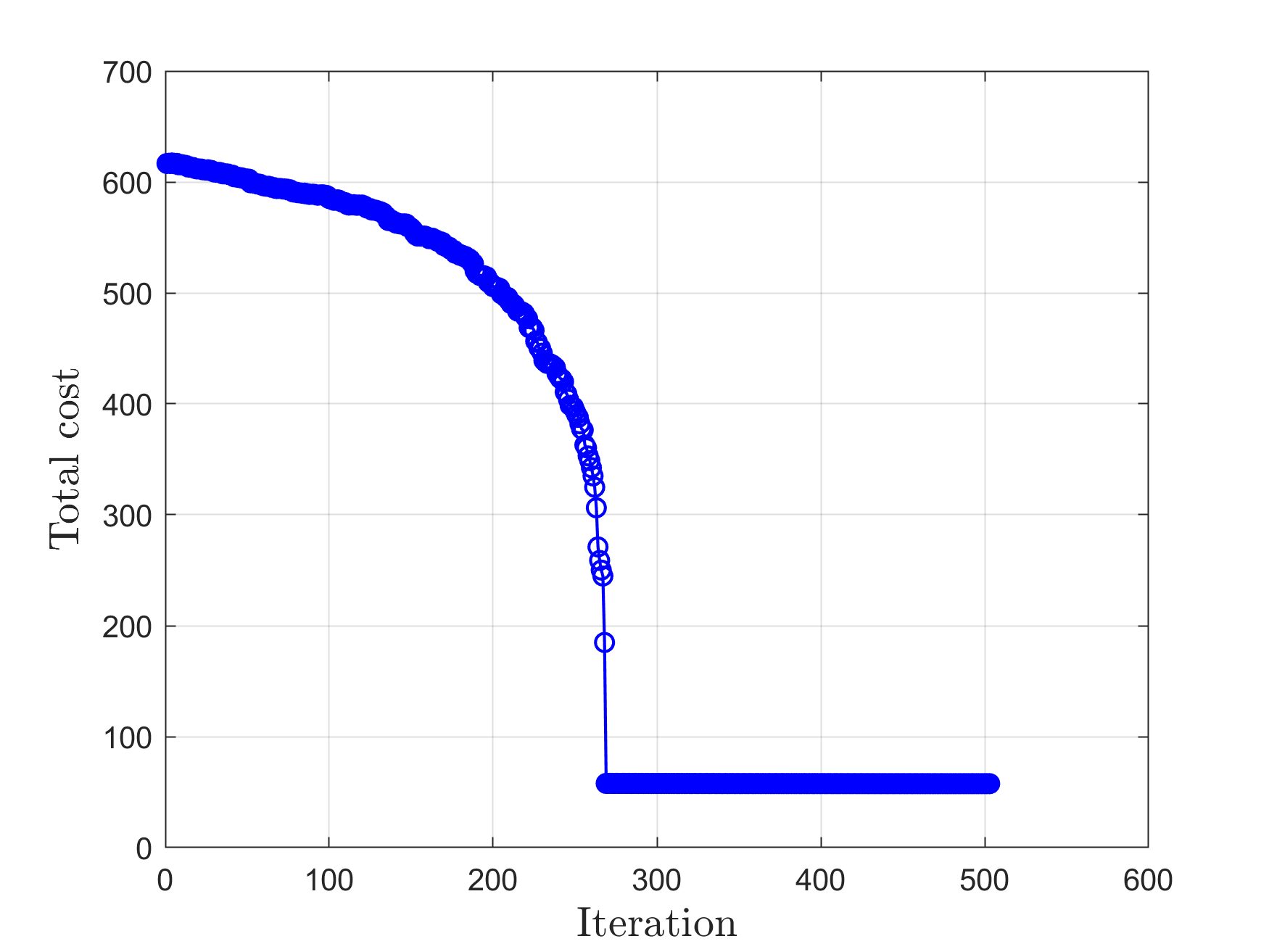}}
	\caption{Cost during training}
	\label{fig:2d_test_cost}
\end{figure}

\begin{figure}
	\centering
	\subfloat[][Trajectory - $\ast$ is the constraint violation location\label{fig:3d_traj1}]
	{\includegraphics[width=.99\textwidth]{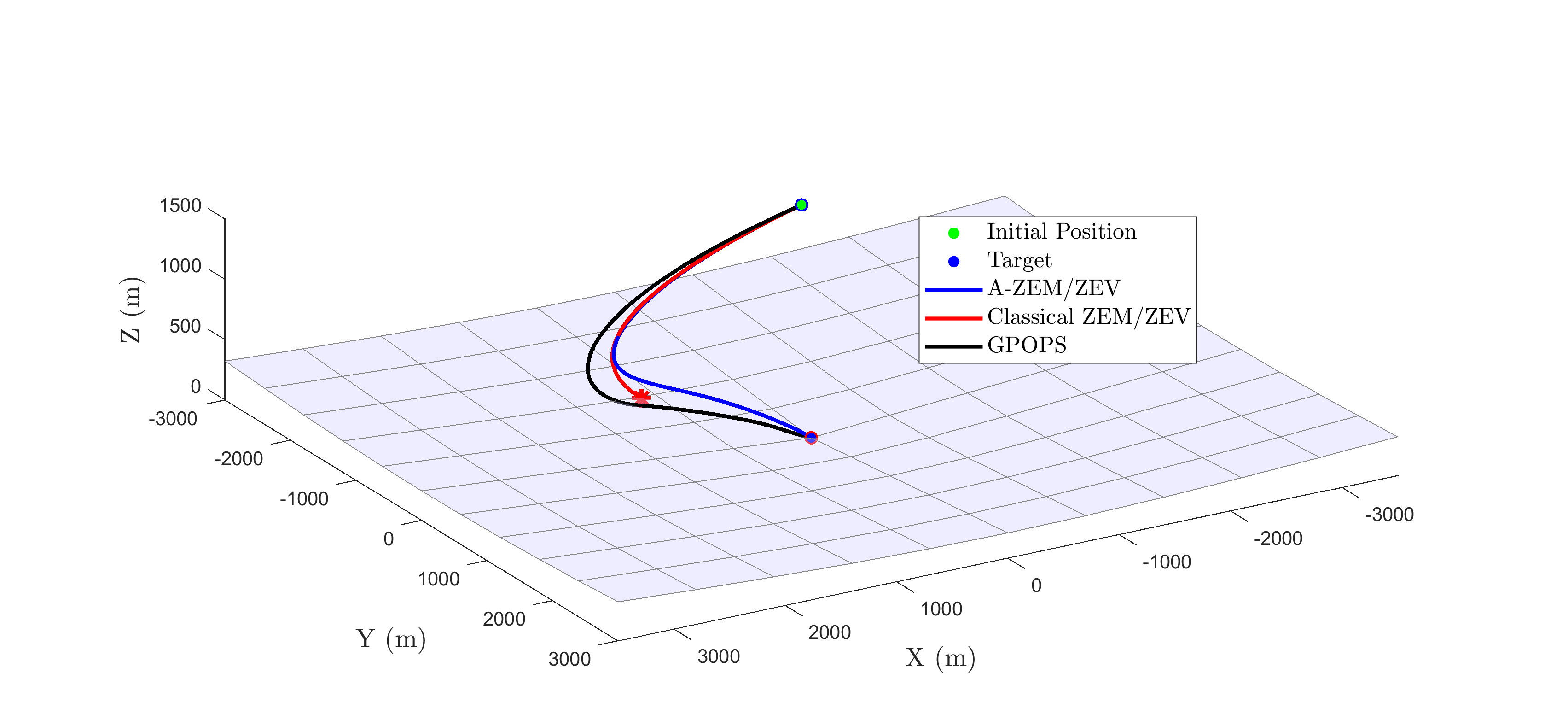}} \\
	\subfloat[][Position\label{fig:3d_pos}]
	{\includegraphics[width=.45\textwidth]{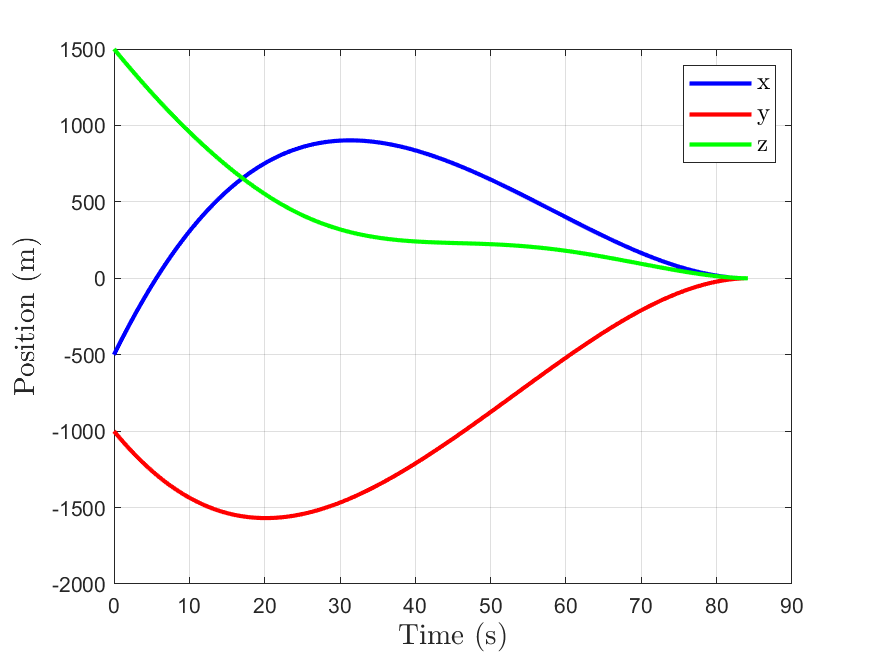}}
	\subfloat[][Velocity\label{fig:3d_vel}]
	{\includegraphics[width=.45\textwidth]{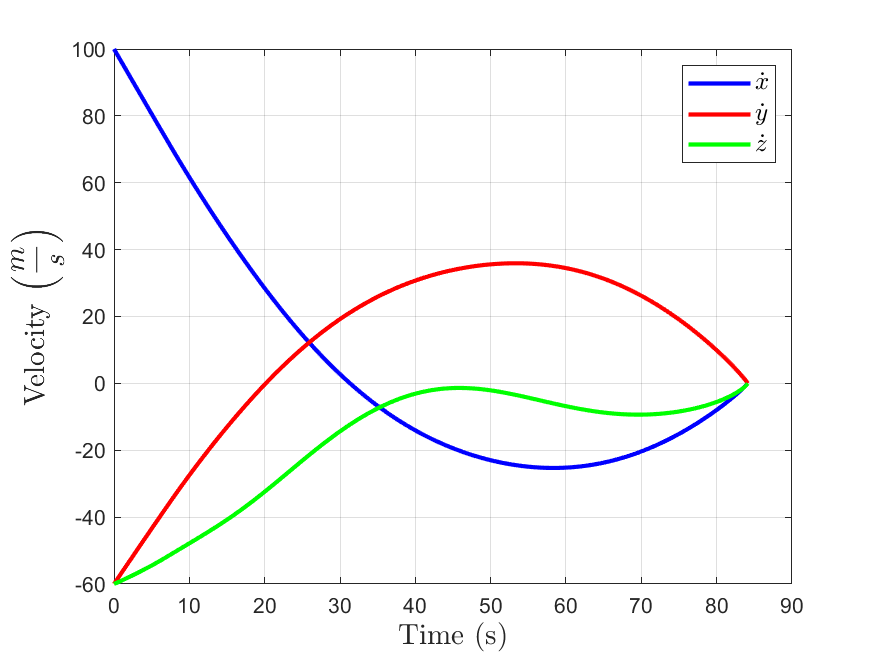}}
	\caption{$ r_0=[-500 , -1000 , 1500]^T m$, $\dot{r_0}=[100 ,-60, -60]^T m/s$}
	\label{fig:3d_traj}
\end{figure}

\begin{figure}
	\centering
	\subfloat[][Thrust\label{fig:3d_thrust}]
	{\includegraphics[width=.45\textwidth]{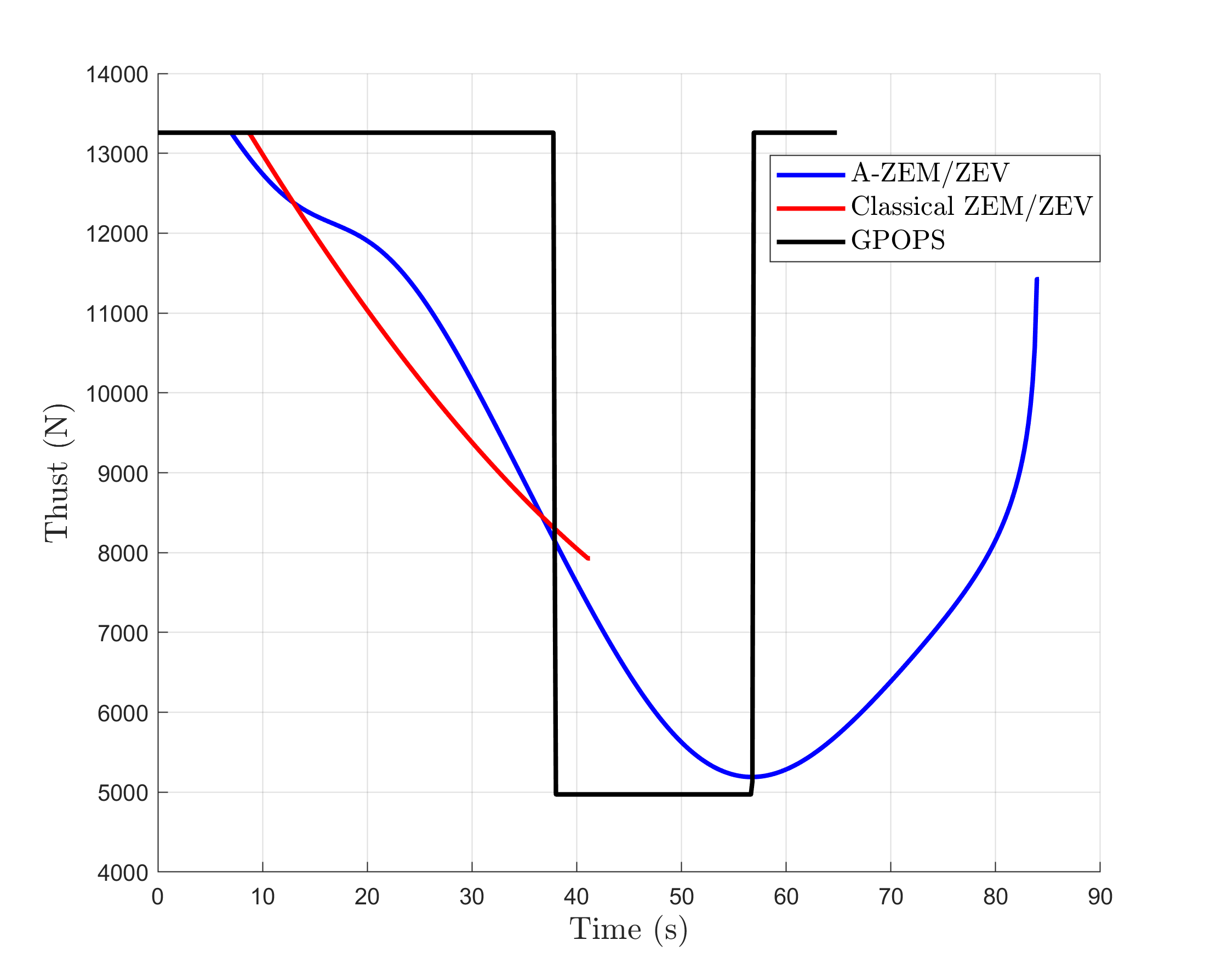}}
	\subfloat[][Spacecraft mass\label{fig:3d_mass}]
	{\includegraphics[width=.45\textwidth]{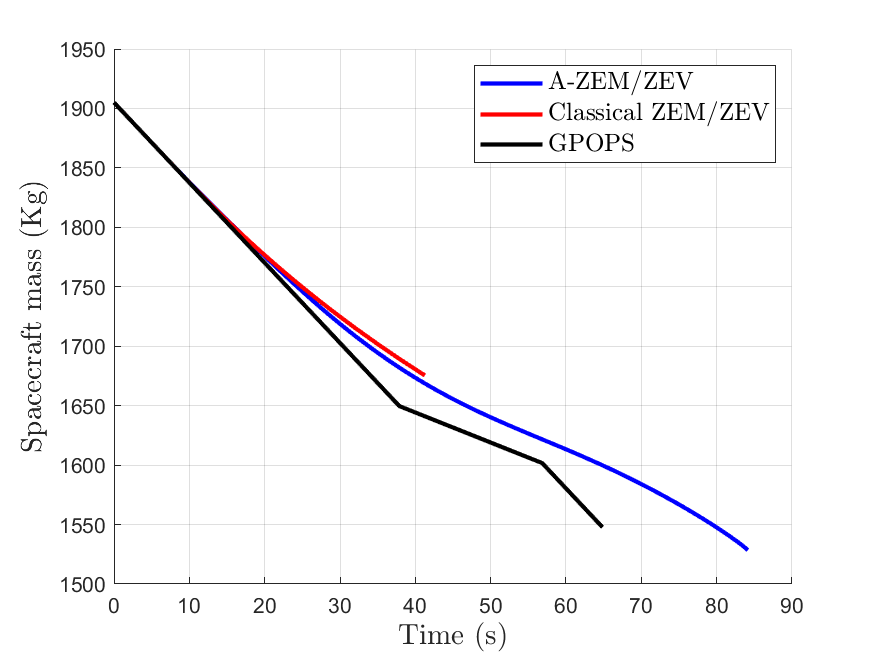}}
	\caption{Thrust, mass and guidance gains for 3D case}
	\label{fig:3d_mass_thrust_gains}
\end{figure}

\begin{figure}
    \centering
    \includegraphics[width=.6\textwidth]{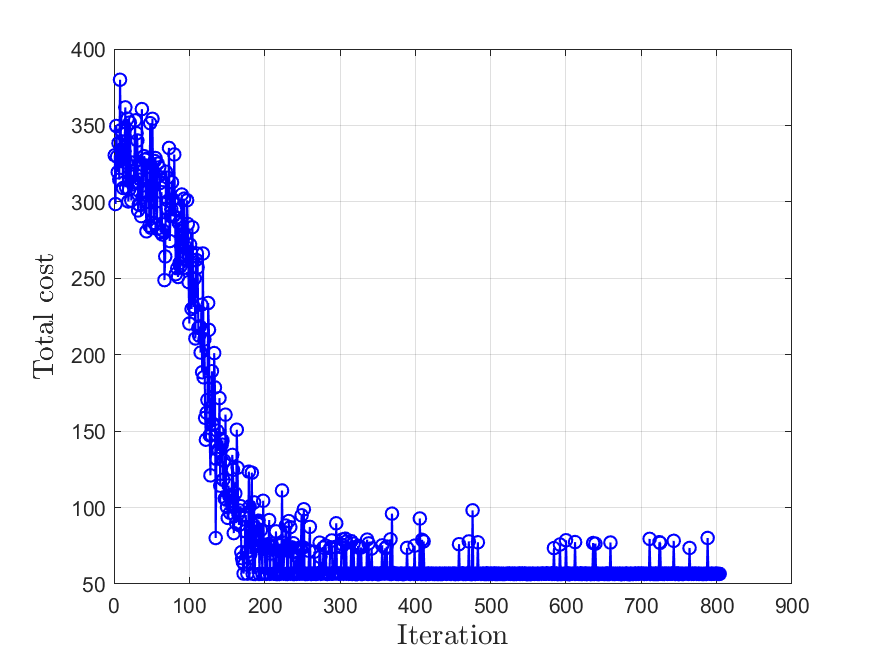}
	\caption{Cost during training}
	\label{fig:3d_cost}
\end{figure}


\begin{figure}
	\centering
	\subfloat[][Guidance gains - 2D \label{fig:gains_2D}]
	{\includegraphics[width=.45\textwidth]{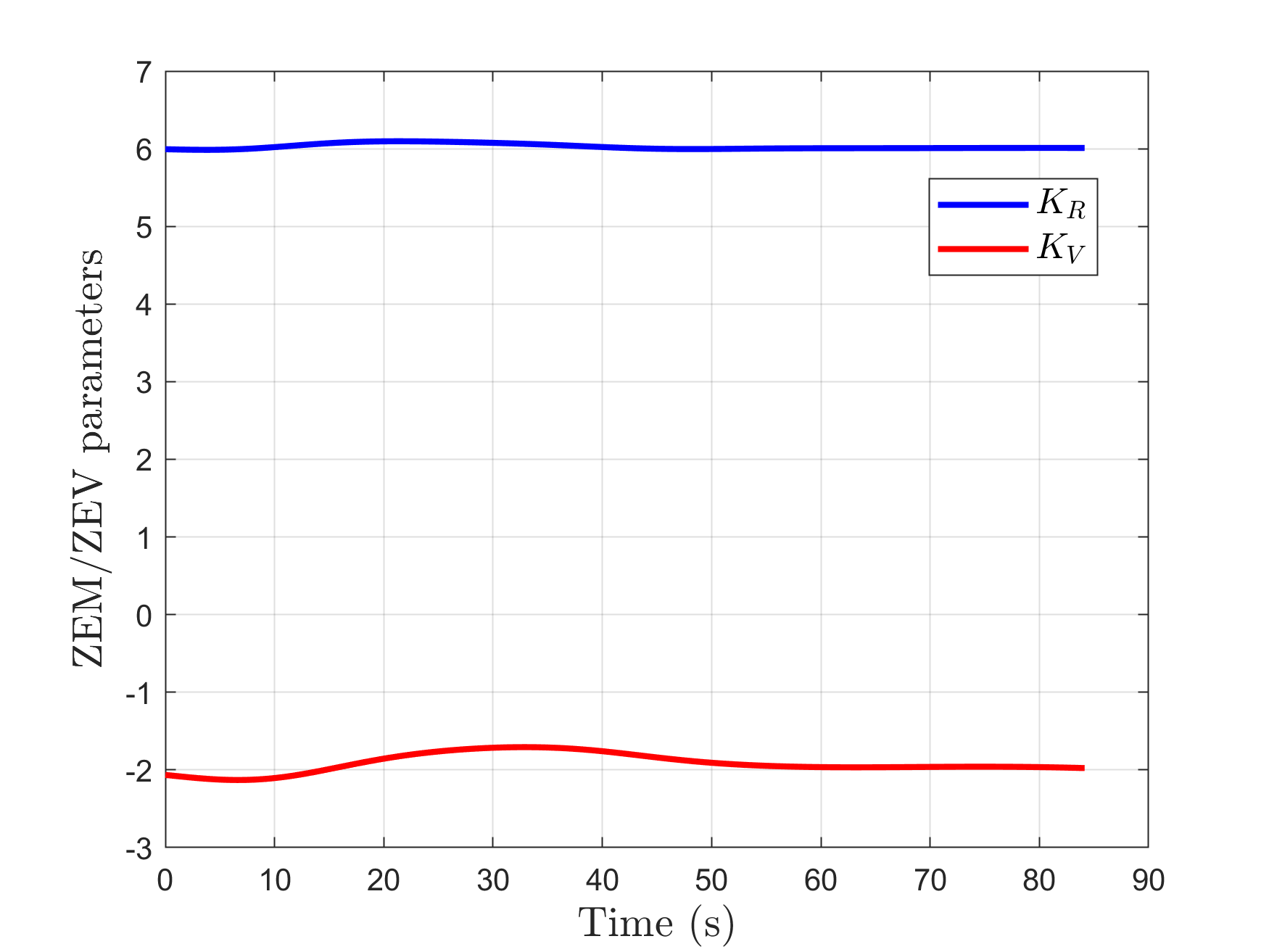}}
	\subfloat[][Guidance gains - 3D \label{fig:gains_3D}]
	{\includegraphics[width=.45\textwidth]{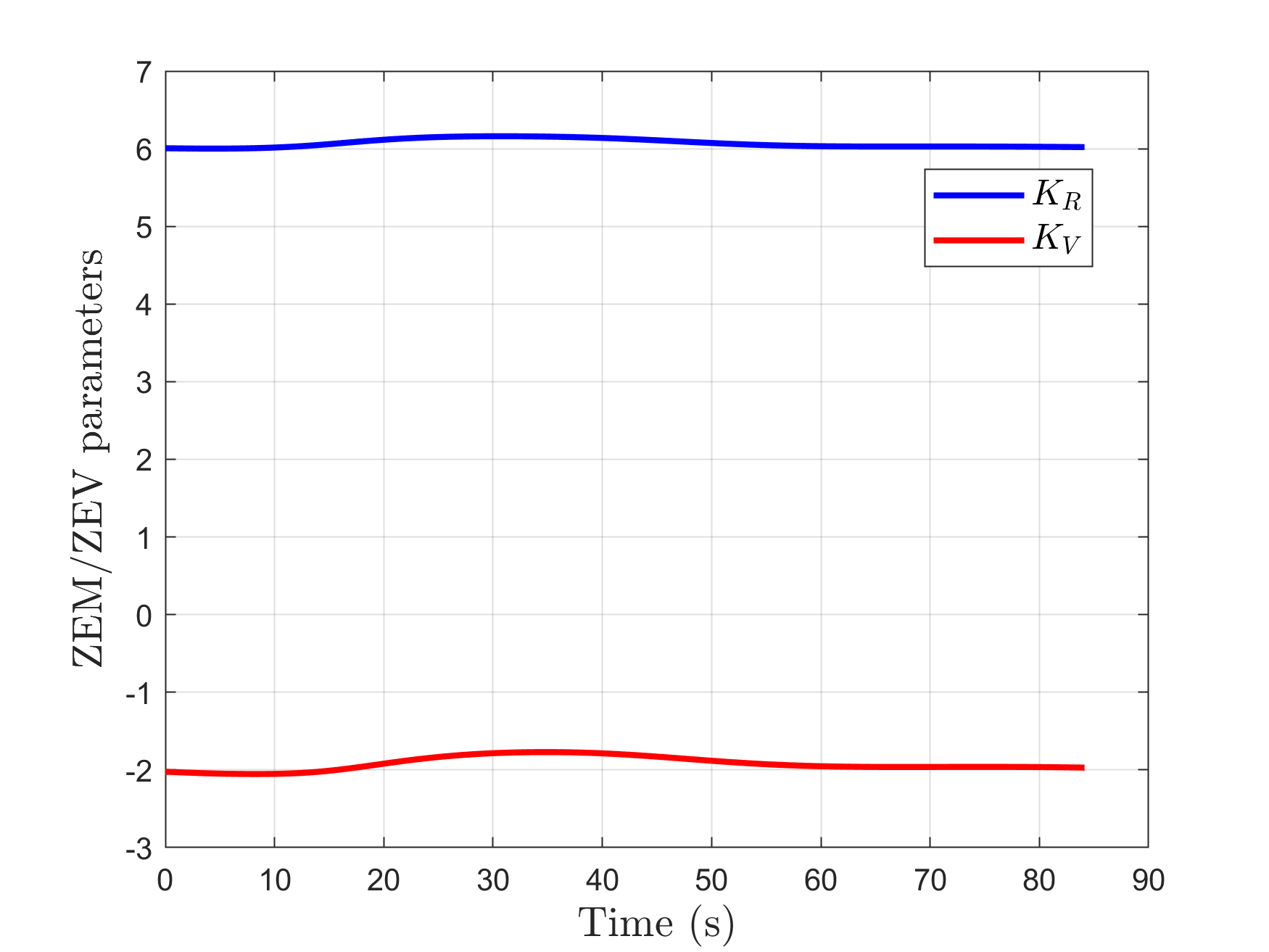}}
	\label{fig:gains}
\end{figure}

The resuls in terms of fuel consumption are shown in Table \ref{tab:performance}. A-ZEM/ZEV performs less optimally with respect to the optimal GPOPS solution as expected but slightly more optimally than Classical-ZEM/ZEV, while managing to avoid collision with the constraint. Classical ZEM/ZEV instead reaches the target but violates the constraint. This shows the major shortcoming of classical ZEM/ZEV: the impossibility to enforce path constraints and why out algorithm overcomes this by making the gains state dependent.

\begin{table}
	\caption{Performance comparison}
	\label{tab:performance}
	\centering
	\begin{tabular}{l | l | c | c} 
		\toprule
		{\bfseries Test case} & {\bfseries Algorithm} & {\bfseries Mass depleted (kg)} & {\bfseries TOF (s)}\\
		\midrule
		\multirow{3}{*}{2D}	    & A-ZEM/ZEV	 				& 382.75 & 84.1\\
								& Classical-ZEM/ZEV		    & 385.51 & 84.1\\ 
								& GPOPS						& 352.59 & 64.7\\ 
		\midrule
		\multirow{3}{*}{3D}     & A-ZEM/ZEV					& 376.54 & 84.1\\
		                        & Classical-ZEM/ZEV		    & 378.81 & 84.1\\ 
		                        & GPOPS						& 357.25 & 64.8\\ 
		\bottomrule
	\end{tabular}
\end{table}

One of the major strengths of the algorithm is the ability to provide a closed-loop guidance control that is both close to optimal and compliant with the constraints. An interesting remark emerges from Figures \ref{fig:gains_2D} and \ref{fig:gains_3D}. Here the evolution of the control gains $K_r$ and $K_v$ is shown. It possible to see that the learning algorithm adjusts the values of the gains according to the constraint scenario in a way that allows the lander to avoid collisions and get to the target safely. The power of the method resides also in the fact that the underlying structure of the ZEM-ZEV guidance allows the algorithm to achieve pinpoint landing accuracy as shown in the following section. The TOF ($T_f$) is optimized by the learning algorithm as a function of the initial state, as explained in section \ref{sec:policy}, and is not modified during the trajectory. The optimal value can be seen in Table \ref{tab:performance}. It should be noted that the TOF for the Classical-ZEM-ZEV is the one obtained after the training process so it has the same value as the one for A-ZEM/ZEV. The TOF of the optimal solution is instead optimized with GPOPS itself.

\subsection{Monte-Carlo analysis}
\label{sec:montecarlo}

\begin{figure}
	\centering
	\subfloat[][Trials\label{fig:mc_trials}]
	{\includegraphics[width=.60\textwidth]{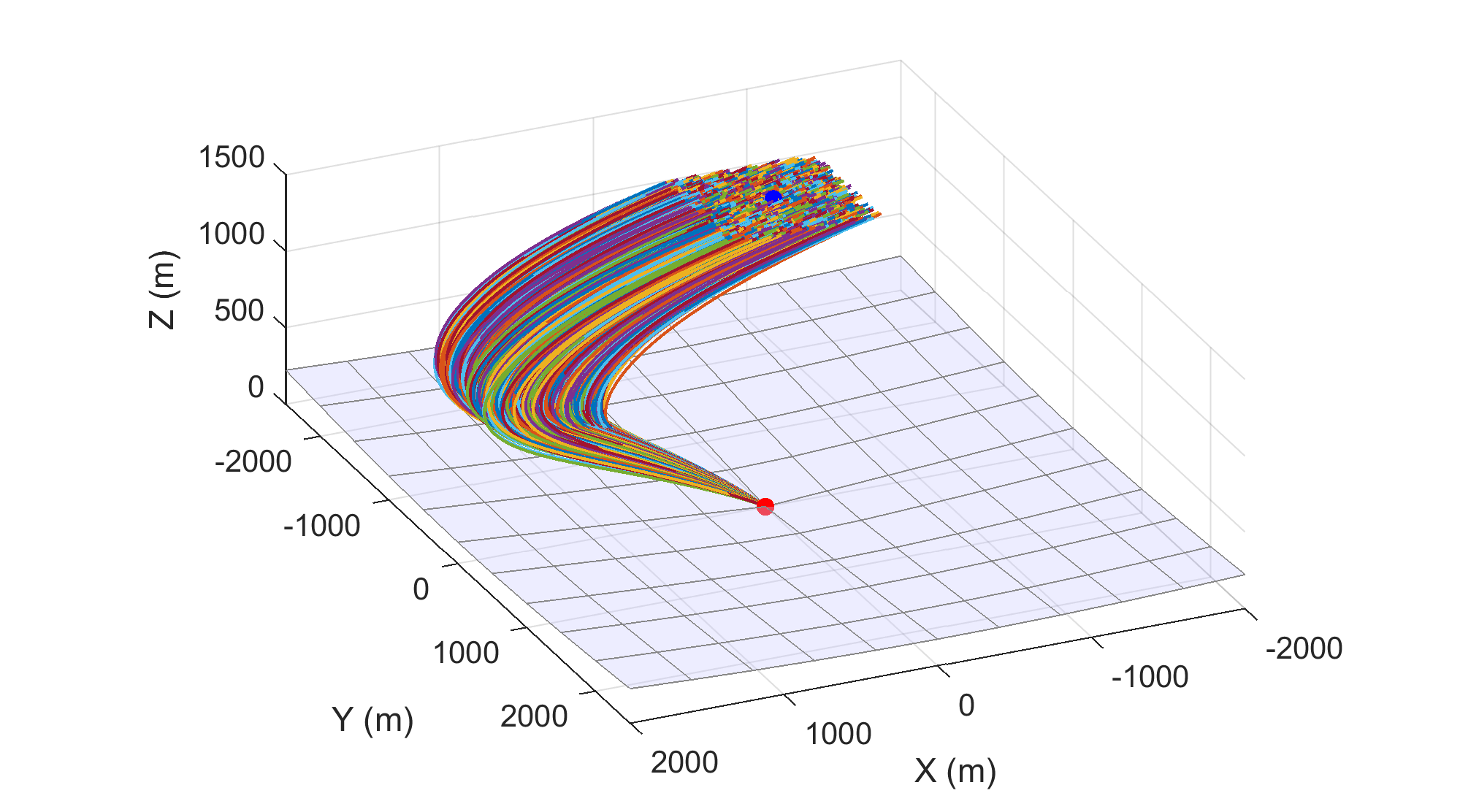}} \\
	\subfloat[][Final position\label{fig:mc_pos}]
	{\includegraphics[width=.48\textwidth]{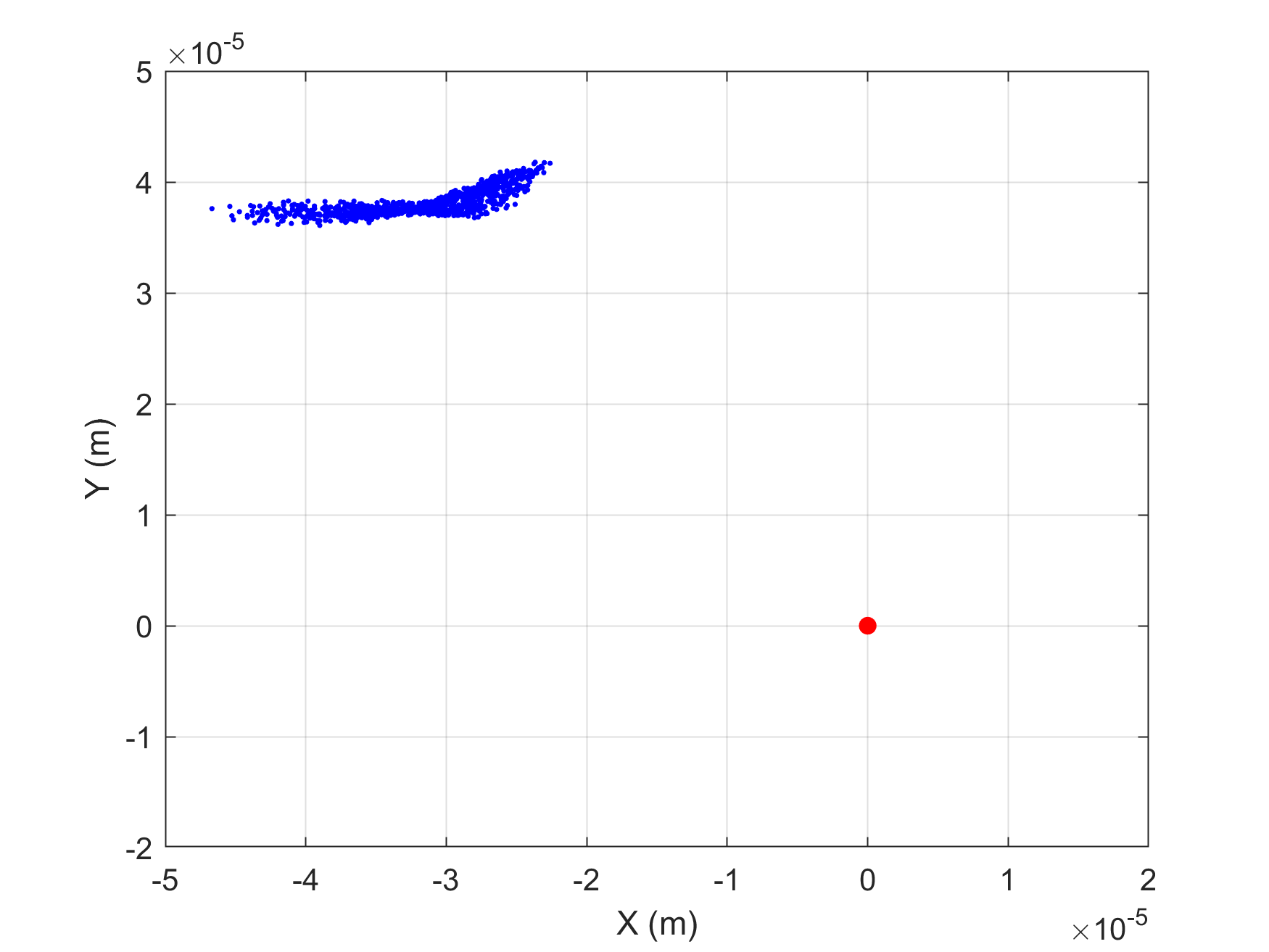}}
	\subfloat[][Final velocity\label{fig:mc_vel}]
	{\includegraphics[width=.48\textwidth]{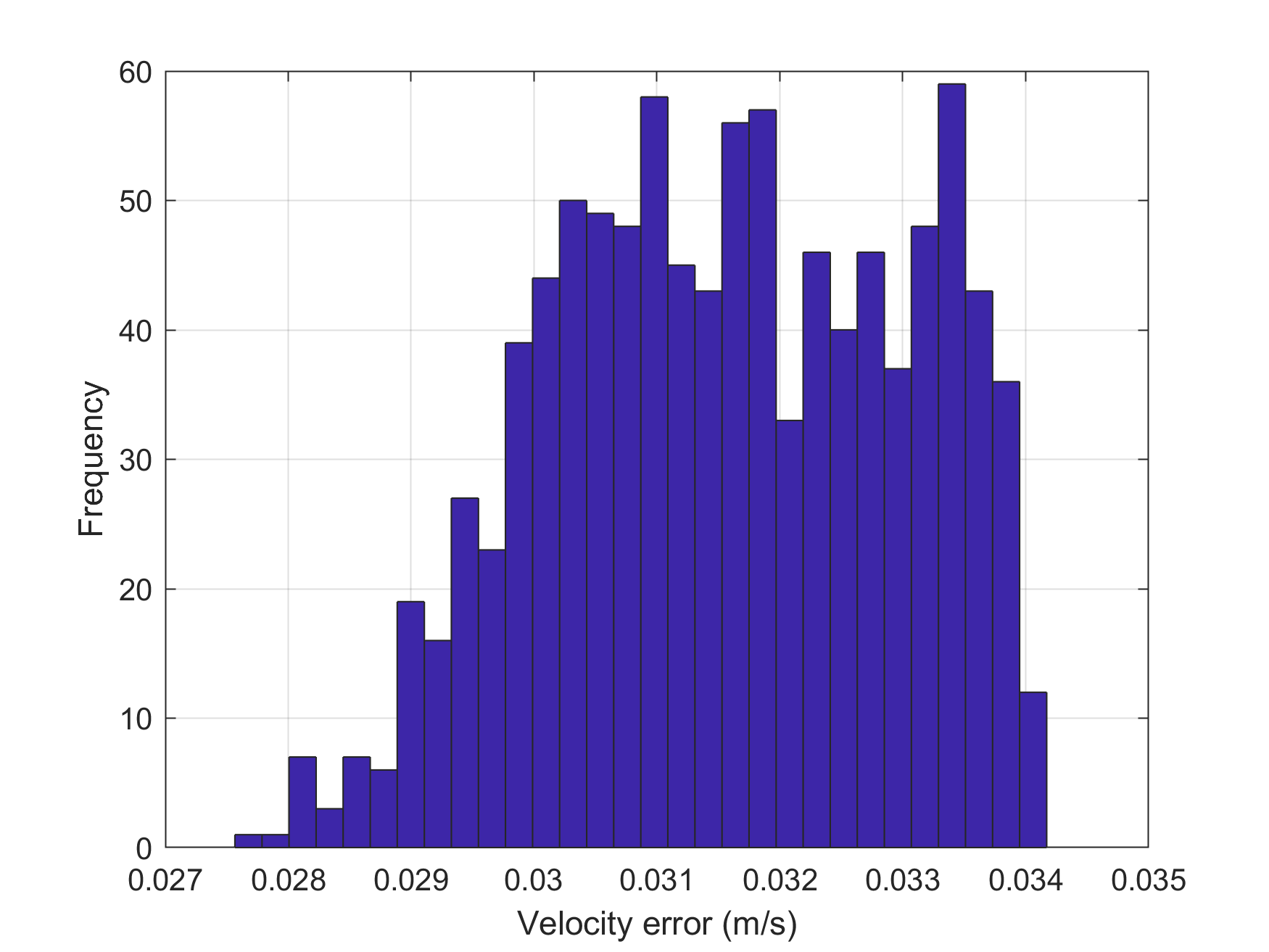}}
	\caption{Monte-Carlo analysis}
	\label{fig:montecarlo}
\end{figure}

A Monte-Carlo analysis was carried out on the 3D case. The objective is to prove that the trained agent is able to perform pinpoint landing with a high degree of accuracy both in terms of final position and velocity. In this case, following the procedure described above, the neural network was trained by selecting the initial state of each sample trajectory from a quite large uniform distribution around the nominal start state. In particular the $x$ and $y$ are taken from a distribution with bounds $\pm 500$ $m$, the $z$ coordinate is kept at $1500$ $m$. The velocity instead has bounds $\pm 5$ $m/s$ . Figure \ref{fig:mc_pos} shows the distribution of the final position on the ground across 1000 trials after training. Figure \ref{fig:mc_vel} shows the distribution of final velocity magnitude. The trained policy clearly manages to drive the spacecraft to the target without ever violating the constraint and with a high degree of accuracy in terms of position, as well as keeping the final velocity below a safe 5 $cm/s$.

\section{Stability Analysis}
\label{chap:stability}

A guidance algorithm should in general be stable so that it can be safely used in practice. In the case of ideal, unperturbed dynamics, it has already been demonstrated that the classical ZEM/ZEV described in Section \ref{sec:Classical_ZEMZEV} is stable \cite{Furfaro:2016:Robust_ZEMZEV}. In this section we study the stability of the A-ZEM/ZEV algorithm.

\subsection{Closed-loop Dynamics}
The formulation of the guidance acceleration as function of ZEM and ZEV results is a linear, non-autonomous, feedback dynamical system. Consequently, classical linear system method of analysis can be employed. The acceleration command for the generalized ZEM/ZEV, as expressed in Section \ref{sec:Classical_ZEMZEV}, is
\begin{equation}
\mathbf{a_c}=\frac{K_R}{t_{go}^2}\mathbf{ZEM} + \frac{K_V}{t_{go}}\mathbf{ZEV}
\end{equation}
considering then that
\begin{align}
\begin{split}
& \dot{\mathbf{ZEM}}=-\mathbf{a_c} t_{go}\\
& \dot{\mathbf{ZEV}}=-\mathbf{a_c} 
\end{split}
\end{align}
the guidance system can be expressed as follows:
\begin{equation}
\label{eq:ZMZV_matrix}
\left[ \begin{matrix}
\dot{\mathbf{ZEM}}\\
\dot{\mathbf{ZEV}} 
\end{matrix} \right] = 
\left[ \begin{matrix}
-\frac{K_R}{t_{go}}  & -K_V  \\
-\frac{K_R}{t_{go}^2}  & -\frac{K_V}{t_{go}} 
\end{matrix} \right] 
\left[ \begin{matrix}
\mathbf{ZEM} \\
\mathbf{ZEV}
\end{matrix} \right]
\end{equation}
In order to study the stability of the Linear Time-Varying (LTV) system one can utilize known properties of linear systems in order to reduce the system to an equivalent Linear Time-Invariant (LTI) system that is much more convenient for stability analysis. 


\subsection{Transformation of LTV systems into LTI systems}
\label{sec:LTV_LTI}
Following the work by Wu \cite{Wu:1978:stability_wu}, let us consider a linear time-varying system
\begin{equation}
\label{eq:ltv}
\dot{\mathbf{x}}=A(t)\mathbf{x}
\end{equation}

\paragraph{\textit{Definition 1}}
A linear time-varying system of the form of Eq.(\ref{eq:ltv}) is said to be \textit{invariable} if it can be transformed into a linear time-invariant system of the form $\dot{\mathbf{z}}=F\mathbf{z}$ by some valid transformations (such as the algegraic transformation and the $t\longleftrightarrow \tau$ transformation defined below), where $F$ is a constant matrix and $\mathbf{z}$ may or may not be an explicit function of $t$.
\paragraph{\textit{Definition 2}}
An \textit{algebraic transformation} is a transformation of states defined by $\textbf{x} (t) = T(t) \bar{\mathbf{x}}(t)$ where $T(t)$ is a non-singular matrix for all $t$ and $\dot{T}(t)$ exists.
\paragraph{\textit{Definition 3}}
A $t\longleftrightarrow \tau$ transformation is a transformation of time scale from $t$ to $\tau$ and is defined by a function of the form $\tau=g(t)$

The invariable systems can be of two different kinds:
\begin{enumerate}
	\item Algebraically invariable systems: the LTV systems that can be transformed into LTI by means of an algebraic transformation alone.
	\item $\tau$-algebraically invariable systems: the LTV systems that can be transformed into LTI systems using an algebraic transformation plus the $t\longleftrightarrow \tau$ transformation.
\end{enumerate}
It can be demonstrated that an LTV of the form Eq.(\ref{eq:ltv}) is invariable \cite{Wu:1978:stability_wu}. It is also algebraically invariable if the state transition matrix (STM) of the system can be found. Unfortunately, in this case, the definition of such STM is extremely cumbersome. Consequently, a $t\longleftrightarrow \tau$ transformation must be employed. The following theorem is valid for $\tau$-algebraically invariable systems.

\paragraph{\textit{Theorem}:}
{\em The linear time-varying system 
\begin{equation}
\label{eq:LTV_2}
\dot{\mathbf{x}}=A(t)\mathbf{x}
\end{equation}
is $\tau$-algebraically invariable if the STM of the system in Eq.~\eqref{eq:LTV_2} is of the form
\begin{equation}
\label{eq:STM_Rg}
\Phi(t,t_0)=T(t,t_0)\exp[Rg(t,t_0)], \quad T(t_0,t_0)=\textbf{I}
\end{equation}
where $\dot{g}(t)$ exists and $t_0$ is chosen so that $g(t_0,t_0)=0$.}

In particular, the algebraic transformation
\begin{equation}
\textbf{x} (t) = T(t,t_0) \bar{\mathbf{x}}(t)
\end{equation}
together with the $t\longleftrightarrow \tau$ transformation
\begin{equation}
\tau=g(t,t_0)
\end{equation}
will transform the system of  Eq.~\eqref{eq:LTV_2} into the time-invariant system 
\begin{equation}
\dot{\mathbf{z}}(\tau)=R\mathbf{z}(\tau)
\end{equation}
where $\textbf{z}(\tau)=\bar{\textbf{x}}(t)$ and $\dot{\textbf{z}}(\tau)=d\textbf{z}(\tau)/d\tau$

Note that by using the definition of $\Phi(t,t_0)$ and the fact that $\text{exp}[Rg(t,t_0)]$ is non-singular, from \ref{eq:STM_Rg} we have:
\begin{equation}
A(t)T(t,t_0)-\dot{T}(t,t_0)=T(t,t_0)R\dot{g}(t,t_0)
\end{equation}
which means
\begin{equation}
\label{eq:RgDot}
R\dot{g}(t,t_0)=T^{-1} \left( A(t)T-\dot{T} \right)
\end{equation}
which will be important in the following section. Once the LTV system has been transformed into a LTI one, the problem of stability is addressed. There are two paths that can be taken in order to prove stability:
\begin{itemize}
	\item The eigenvalues of the LTI system matrix $R$ are computed; if the real part of all the eigenvalues is negative or at most 0, the system is stable.
	\item The State Transition Matrix (STM) of the original LTV system is retrieved. If it is possible to demonstrate that it is bounded at all time, then the system is stable.
\end{itemize}


\subsection{Stability of the A-ZEM/ZEV algorithm}
\label{sec:stability_AZEMZEV}
 For the A-ZEM/ZEV guidance, the matrix $A$ is the following:
\begin{equation}
\label{eq:matrix_generalize_ZEMZEV}
A=\left[ \begin{matrix}
-\frac{K_R}{t_{go}}  & -K_V \\
-\frac{K_R}{t_{go}^2}  & -\frac{K_V}{t_{go}} 
\end{matrix} \right]
\end{equation}
using Eq.(\ref{eq:RgDot}) with
\begin{equation}
T=\left[ \begin{matrix}
1  & 0  \\
0  & \frac{t_f}{t_{go}} 
\end{matrix} \right]
\end{equation}
the system in Eq.(\ref{eq:matrix_generalize_ZEMZEV}) can be transformed in the algebraically equivalent system, as follows:
\begin{equation}
\label{eq:algebraically_equivalent}
R\dot{g}(t,t_0)=
\left[ \begin{matrix}
-K_R  & -K_V t_f \\
-\frac{K_R}{t_{f}}  & -(K_V + 1)
\end{matrix} \right] \frac{1}{t_{go}}
\end{equation}
with
\begin{equation}
\label{eq:Rmatrix}
R=
\left[ \begin{matrix}
-K_R  & -K_V t_f \\
-\frac{K_R}{t_{f}}  & -(K_V + 1)
\end{matrix} \right] 
\end{equation}
and
\begin{equation}
\dot{g}(t,t_0)=\frac{1}{t_{go}}
\end{equation}
The resulting system is simpler but still dependent on time. In order to make it time-invariant, the $t\longleftrightarrow \tau$ transformation must be applied. The time basis transformation is
\begin{equation}
\tau=g(t,t_0)=\int_{t_0}^{t} \frac{1}{t_{go}} d\tau=-\log \frac{t_{go}}{t_f}
\end{equation}
where $t_0$ has been chosen so that $g(t_0,t_0)=0$. With this transformation, the system is now a LTI system
\begin{equation}
\dot{\mathbf{z}}(\tau)=R\mathbf{z}(\tau)
\end{equation}
with system matrix $R$. The stability of the system can be proven by finding the eigenvalues of such matrix and prove they have negative real part at all times. The $R$ matrix in Eq.(\ref{eq:Rmatrix}) has eigenvalues
\begin{equation}
\lambda_{1,2}=\frac{-K \pm \sqrt{K^2-4K_R}}{2}, \quad K=K_R+K_V+1
\end{equation}
The stability conditions can be found in two cases: $\Delta \geq 0$ and $\Delta < 0$, where $\Delta=K^2-4K_R$. 

\paragraph{Case 1: When $\mathbf{\Delta \geq 0}$. \\}
The condition $\Delta \geq 0$ translates into
\begin{equation}
\label{eq:delta_cond}
K_V^2+K_R^2+2 K_R K_V +2 K_V - 2 K_R +1 \geq 0
\end{equation}
and means that the eigenvalues are purely real. \ref{eq:delta_cond} must be verified in order for the following stability condition to hold:
\begin{equation}
-K \pm \sqrt{\Delta} < 0
\end{equation}
or
\begin{align}
\begin{split}
& -K + \sqrt{\Delta} < 0 \quad \rightarrow \quad K>\sqrt{\Delta} \\
& -K - \sqrt{\Delta} < 0 \quad \rightarrow \quad K>-\sqrt{\Delta}
\end{split}
\end{align}
which means that, since $\sqrt{\Delta}>0$, the condition for stability is
\begin{equation}
\label{eq:stab_cond_1}
K>\sqrt{\Delta} = \sqrt{K^2-4K_R}
\end{equation}

\begin{figure}
	\centering
	\subfloat[][Case 1: 2D\label{fig:2d_eig}]
	{\includegraphics[width=.45\textwidth]{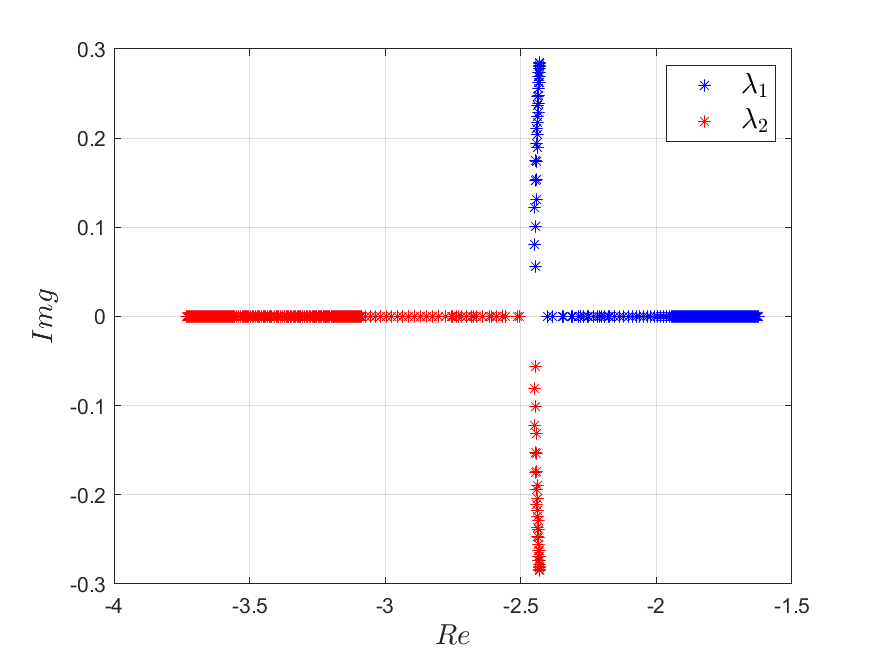}}
	\subfloat[][Case 2: 3D\label{fig:3d_eig}]
	{\includegraphics[width=.45\textwidth]{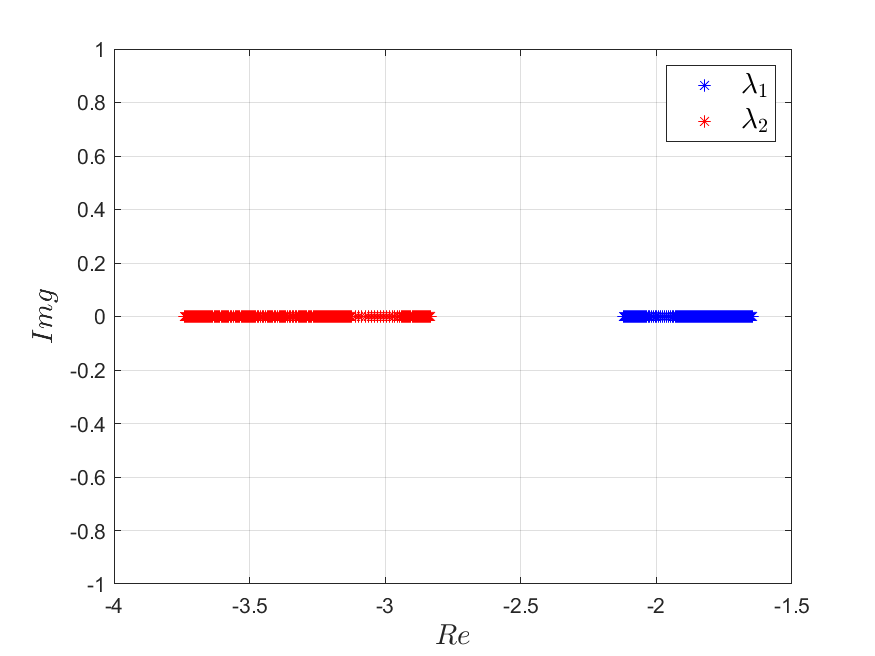}}
	\caption{Eigenvalues during A-ZEM/ZEV-guided powered descent}
	\label{fig:eig}
\end{figure}

\paragraph{Case 2: When $\mathbf{\Delta<0}$. \\}
If $\Delta<0$, so
\begin{equation}
K_V^2+K_R^2+2 K_R K_V +2 K_V - 2 K_R +1 < 0
\end{equation}
the eigenvalues have a real and an imaginary part. In order for the system to be stable, the real part must be negative. So in this case the stability condition is simply
\begin{equation}
\label{eq:stab_cond_2}
-K < 0 \quad \rightarrow \quad K>0 \quad \rightarrow \quad K_R+K_V+1>0
\end{equation}
These two conditions offer a quick way to check for the instantaneous stability of the guidance algorithm. This can be added as a checking step inside the control loop with a relatively low computational cost and action can be taken in case the algorithm goes in unstable regions. To prove that the algorithm remains stable in our test cases, the eigenvalues were computed along the descent trajectories in both cases (both 2D and 3D) and the results are reported in Figure \ref{fig:eig}. It is clear that the real part of the eigenvalues all remains strictly negative which ensures stability.

\begin{figure}
	\centering
	\subfloat[][Case 1: 2D\label{fig:2d_stm}]
	{\includegraphics[width=.45\textwidth]{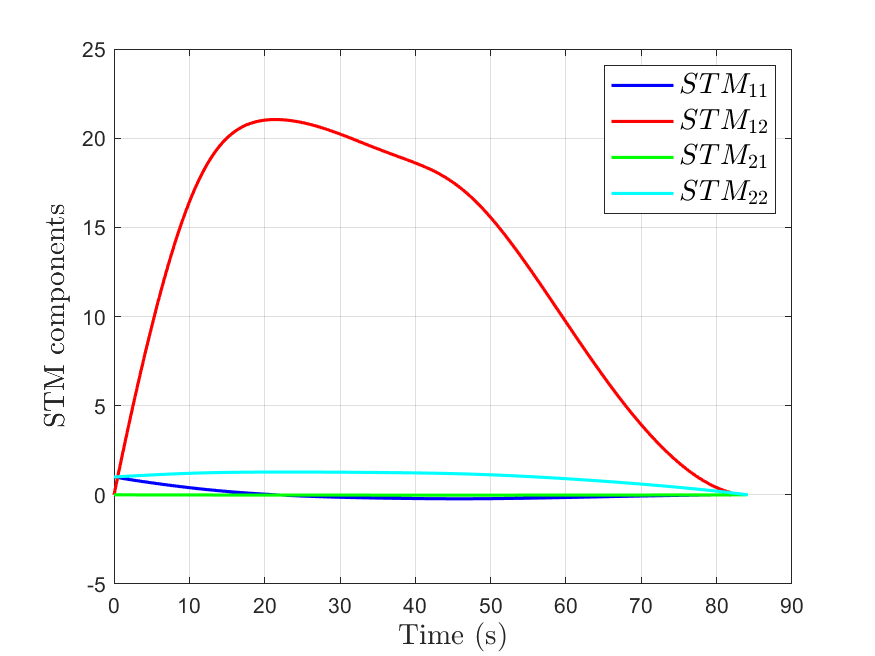}}
	\subfloat[][Case 2: 3D\label{fig:3d_stm}]
	{\includegraphics[width=.45\textwidth]{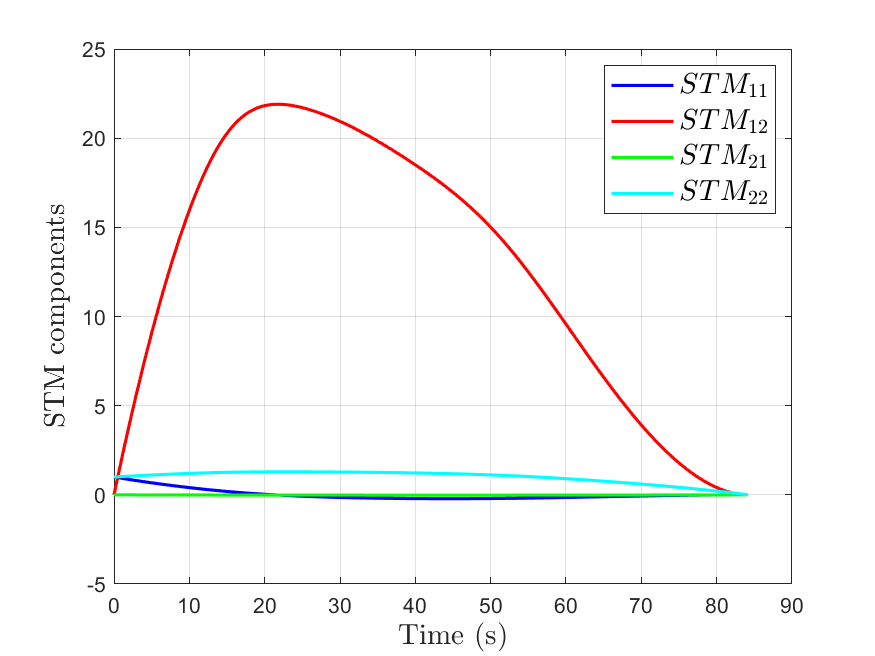}}
	\caption{State transition matrix components}
	\label{fig:STM}
\end{figure}

As stated in Section \ref{sec:LTV_LTI}, another way of addressing the stability of the closed-loop dynamics is to employ the State Transition Matrix (STM).  In particular, the STM of the LTV system must be bounded at all times in order for it to be stable. The calculation of the state transition matrix of the LTV system is derived from the STM of the LTI system. Letting $\Phi^*$ be the STM of the LTI system, then the STM of the original LTV system is $\Phi(t,t_0)=T\Phi^*$.
According to \cite{Rowell:2002:STM}, the STM of a linear system can be found from the knowledge of eigenvalues and eigenvectors as
\begin{equation}
\Phi(t,t_0)=\textbf{M}e^{\mathbf{\Lambda} t}\textbf{M}^{-1}
\end{equation}
Where
\begin{equation}
\textbf{M}=\left[ \begin{matrix}
\textbf{m}_1  \quad &| \quad \textbf{m}_2 \quad &| \quad \dots  \quad &| \quad \textbf{m}_n
\end{matrix} \right]
\end{equation}
is the matrix whose columns are the eigenvectors and
\begin{equation}
\mathbf{\Lambda}=
\left[ \begin{matrix}
\lambda_1  & 0 & \dots & 0\\
0 & \lambda_2 & \dots & 0 \\
\vdots & \vdots & \ddots & \vdots \\
0 & 0 & \dots & \lambda_n
\end{matrix} \right]
\end{equation}
is the diagonal matrix of the eigenvalues.
In this case, applying such definitions to the algebraically equivalent system in \ref{eq:algebraically_equivalent} and then the $T$ transformation, the STM of the original system turns out to be:

\begin{equation}
\label{eq:STM_original}
\Phi= \begin{bmatrix}
\Phi_{11}
& \Phi_{12} \\
\Phi_{21} & \Phi_{22}
\end{bmatrix}
\end{equation}

with
\begin{eqnarray}
\label{eq:STM_component}
& \Phi_{11} =  
\frac{\lambda_1+k_r}{\lambda_1-\lambda_2} \left(  \frac{t_{go}}{t_f} \right)^{-\lambda_2}
-\frac{\lambda_2+k_r}{\lambda_1-\lambda_2} \left(  \frac{t_{go}}{t_f} \right)^{-\lambda_1} & \\
& \Phi_{12} = -\frac{k_v t_f}{\lambda_1-\lambda_2} \left(  \frac{t_{go}}{t_f} \right)^{-\lambda_1} + 
\frac{k_v t_f}{\lambda_1-\lambda_2} \left(  \frac{t_{go}}{t_f} \right)^{-\lambda_2} & \\
& \Phi_{21} = \frac{\lambda_1+k_r}{k_v t_f} \frac{\lambda_2+k_r}{\lambda_1-\lambda_2} \left(  \frac{t_{go}}{t_f} \right)^{-\lambda_1-1} -
\frac{\lambda_2+k_r}{k_v t_f} \frac{\lambda_1+k_r}{\lambda_1-\lambda_2} \left(  \frac{t_{go}}{t_f} \right)^{-\lambda_2-1} & \\
& \Phi_{22} = \frac{\lambda_1+k_r}{\lambda_1-\lambda_2} \left(  \frac{t_{go}}{t_f} \right)^{-\lambda_1-1} - 
\frac{\lambda_2+k_r}{\lambda_1-\lambda_2} \left(  \frac{t_{go}}{t_f} \right)^{-\lambda_2-1} &
\end{eqnarray}


where
\begin{equation}
\lambda_{1,2}=\frac{-K \pm \sqrt{K^2-4K_R}}{2}, \quad K=K_R+K_V+1
\end{equation}
According to the theory on non-autonomous linear systems, the LTV in equation \ref{eq:ltv} is stable if and only if the STM in \ref{eq:STM_original} is bounded at all time $0\leq t \leq t_f$. The STM in equation \ref{eq:STM_original} was computed along the entire guided descent trajectories for both 2D and 3D cases. Figure \ref{fig:STM} shows the STM for two sampled guided trajectories. It is clear that the components of the STM are bounded at all times so we can conclude that, at least in these cases, the algorithm is stable.

\section{Conclusions and future efforts} \label{sec:conclusion}
The work has shown a novel closed-loop spacecraft guidance algorithm for soft landing. Using machine learning, in particular an actor-critic algorithm based on policy gradient, with advantage function estimation, it has been possible to expand the capabilities of classical ZEM/ZEV feedback guidance. The resulting \textit{adaptive} algorithm (A-ZEM/ZEV) improves its performance in terms of fuel consumption and allows path constraints to be implemented directly into the guidance law. The actor-critic algorithm outputs a trained guidance neural network that can be implemented directly on-board. Provided that the spacecraft is equipped with the appropriate set of sensor for state determination, the guidance is recovered as function of the state in real time and can be followed by the control system, allowing for a completely autonomous soft landing maneuver in presence of path constraints. Moreover the stability of the method has been addressed: by transforming the LTV system in a LTI one, it has been possible to easily check for instability condition, either during post processing or, more importantly, online in the control loop. This ensures that countermeasures can be applied if the unstable regions are visited.

From a machine learning prospective, ELM have shown to work well as critic in an actor-critic algorithm. This shows that when the task is a simple regression problem like in this case, deep learning is not needed. A shallow network is enough and the one-step learning process of ELM makes them perfect for the task. They scale well with input dimension achieving good accuracy with a negligible computational time if compared with the total iteration time.

The work presented in this paper shows the advantages of using machine learning to learn state dependent parameters in an existing parametrized policy. In this case, using ZEM/ZEV guidance as a baseline allows for extremely precise terminal guidance with the increased flexibility given by the usage of reinforcement learning. This can be expanded to other guidance problem that rely on a parametrized law. As long as the state space can be discretized effectively, this ELM-based actor-critic algorithm can be applied. Convergence of the method is guaranteed by the fact that the advantage function estimation is unbiased and the cost function is set up correctly but convergence is still slow. Work can be done to improve the learning performance. For example, meta-learning could be used to learn different sequential task in order to speed up the process, especially if the environmental condition change (i.e. actuator or sensor failure). Meta-learning could also be used to create an algorithm that is more robust mis-modelled dynamics. By learning over a distributions of MDPs corresponding to different randomized instances of the environment, the agent should in fact be able to adapt to a wider set of environmental parameters, ultimately improving the performance in a more realistic set up.


\bibliographystyle{elsarticle-num}

\begin{thebibliography}{999}

\bibitem[]{shotwell2005phoenix} Shotwell, Robert, ``Phoenix—the first Mars Scout mission'' \emph{Acta Astronautica}, Vol. 57, No. 2-8, pp. 121-134, 2005.

\bibitem[]{grotzinger2012mars} Grotzinger, John P and Crisp, Joy and Vasavada, Ashwin R and Anderson, Robert C and Baker, Charles J and Barry, Robert and Blake, David F and Conrad, Pamela and Edgett, Kenneth S and Ferdowski, Bobak , ``Mars Science Laboratory mission and science investigation'' \emph{Space science reviews}, Vol. 170, No. 1-4, pp. 5-56, 2012.

\bibitem[]{burns2018science} Burns, Jack O and Mellinkoff, Benjamin and Spydell, Matthew and Fong, Terrence and Kring, David A and Pratt, William D and Cichan, Timothy and Edwards, Christine M , ``Science on the lunar surface facilitated by low latency telerobotics from a Lunar Orbital Platform-Gateway'' \emph{Acta Astronautica}, 2018.

\bibitem[]{steltzner2010mars} Steltzner, Adam D and Burkhart, P Dan and Chen, Allen and Comeaux, Keith A and Guernsey, Carl S and Kipp, Devin M and Lorenzoni, Leila V and Mendeck, Gavin F and Powell, Richard W and Rivellini, Tommaso P and others , ``Mars science laboratory entry, descent, and landing system overview'' \emph{Pasadena, CA: Jet Propulsion Laboratory, National Aeronautics and Space Administration}, 2010.

\bibitem[]{klumpp1974apollo} Klumpp, Allan R , ``Apollo lunar descent guidance'' \emph{Automatica}, Vol. 10, No. 2, pp. 133-146, 1974.

\bibitem[]{singh2007guidance} Singh, Gurkirpal and SanMartin, Alejandro M and Wong, Edward C , ``Guidance and control design for powered descent and landing on Mars'' \emph{Aerospace Conference, 2007 IEEE}, pp. 1-8, 2007.

\bibitem[A{\c{c}}{\i}kme{\c{s}}e(2007)]{acikmese2007convex} Acikmese, B. and Ploen, S.R., ``Convex programming approach to powered descent guidance for mars landing'' \emph{Journal of Guidance, Control, and Dynamics}, Vol. 30, No. 5, pp. 1353-1366, 2007.

\bibitem[]{blackmore2010minimum} Blackmore, Lars and Acikmese, Behcet and Scharf, Daniel P , ``Minimum-landing-error powered-descent guidance for Mars landing using convex optimization'' \emph{Journal of guidance, control, and dynamics}, Vol. 33, No. 4, pp. 1161-1171, 2010.

\bibitem[]{accikmecse2013lossless} A{\c{c}}{\i}kme{\c{s}}e, Beh{\c{c}}et and Carson, John M and Blackmore, Lars , ``Lossless convexification of nonconvex control bound and pointing constraints of the soft landing optimal control problem'' \emph{IEEE Transactions on Control Systems Technology}, Vol. 21, No. 6, pp. 2104-2113, 2013.

\bibitem[]{trawny2015flight} Trawny, Nikolas and Benito, Joel and Tweddle, Brent E and Bergh, Charles F and Khanoyan, Garen and Vaughan, Geoffrey and Zheng, Jason and Villalpando, Carlos and Cheng, Yang and Scharf, Daniel P and others , ``Flight testing of terrain-relative navigation and large-divert guidance on a VTVL rocket'' \emph{AIAA SPACE 2015 Conference and Exposition}, pp. 4418, 2015.

\bibitem[]{liu2017survey} Liu, Xinfu and Lu, Ping and Pan, Binfeng , ``Survey of convex optimization for aerospace applications'' \emph{Astrodynamics}, Vol. 1, No. 1, pp. 23-40, 2017.

\bibitem[]{lu2017propellant} Lu, P., ``Propellant-Optimal Powered Descent Guidance'' \emph{Journal of Guidance, Control, and Dynamics}, Vol. 41, No. 4, pp. 813-826, 2017.

\bibitem[]{lu2018adaptive} Lu, Ping and Sostaric, Ronald R and Mendeck, Gavin F , ``Adaptive Powered Descent Initiation and Fuel-Optimal Guidance for Mars Applications'' \emph{2018 AIAA Guidance, Navigation, and Control Conference},  pp. 0616, 2018.


\bibitem[]{lu2010highly} Lu, Ping and Pan, Binfeng , ``Highly constrained optimal launch ascent guidance'' \emph{Journal of Guidance, Control, and Dynamics}, Vol. 33, No. 2, pp. 404-414, 2010.

\bibitem[]{lu2008rapid} Lu, Ping and Griffin, Brian J and Dukeman, Gregory A and Chavez, Frank R , ``Rapid optimal multiburn ascent planning and guidance'' \emph{Journal of Guidance, Control, and Dynamics}, Vol. 31, No. 6, pp. 1656-1664, 2008.

\bibitem[]{lu2012versatile} Lu, Ping and Forbes, Stephen and Baldwin, Morgan , ``A versatile powered guidance algorithm'' \emph{AIAA Guidance, Navigation, and Control Conference}, pp. 4843, 2012.

\bibitem[]{guo2013waypoint} Guo, Yanning and Hawkins, Matt and Wie, Bong , ``Waypoint-optimized zero-effort-miss/zero-effort-velocity feedback guidance for mars landing'' \emph{Journal of Guidance, Control, and Dynamics}, Vol. 36, No. 3, pp. 799-809, 2013.


\bibitem[Pinson(2016)]{pinson2016trajectory} Pinson, R. and Lu, P. , ``Trajectory design employing convex optimization for landing on irregularly shaped asteroids'', AIAA/AAS Astrodynamics Specialist Conference, 2016.


\bibitem[Zhang(2018)]{zhang2015low} Zhang, C. and Topputo, F. and Bernelli-Zazzera, F. and Zhao, Y., ``Low-thrust minimum-fuel optimization in the circular restricted three-body problem'' \emph{Journal of Guidance, Control, and Dynamics}, Vol. 38, No. 8, pp. 1501--1510, 2018.

\bibitem[Lu(2013)]{lu2013autonomous} Lu, P. and Liu, X., ``Autonomous trajectory planning for rendezvous and proximity operations by conic optimization'' \emph{Journal of Guidance, Control, and Dynamics}, Vol. 36, No. 2, pp. 375-389, 2013.


\bibitem[Rao(2009)]{rao2009survey} Rao, A.V, ``A survey of numerical methods for optimal control'' \emph{Advances in the Astronautical Sciences}, Vol. 135, No. 1, pp. 497-528, 2009.


\bibitem[Betts(2010)]{betts2010practical} Betts, J.T, ``Practical methods for optimal control and estimation using nonlinear programming'' \emph{Siam}, Vol. 19, 2010.


\bibitem[]{Guo:2013:ZMZV_generalized} Guo, Y. and Hawkins, M. and Wie, B. , ``Applications of generalized zero-effort-miss/zero-effort-velocity feedback guidance algorithm'' \emph{Journal of Guidance, Control, and Dynamics}, Vol. 36, No. 3, pp. 810-820, 2013.

\bibitem[]{Guo:2011:ZMZV_plan_landing_asteroid_inter} Guo, Y. and Hawkins, M. and Wie, B. , ``Optimal feedback guidance algorithms for planetary landing and asteroid intercept'' \emph{AAS/AIAA astrodynamics specialist conference}, Vol. 36, No. 3, pp. 588, 2011.

\bibitem[]{Furfaro:2017:waypoints} Furfaro, R. and Linares, R. , ``Waypoint-Based Generalized ZEM/ZEV Feedback Guidance for Planetary Landing via a Reinforcement Learning Approach'' \emph{3rd IAA Conference on Dynamics and Control of Space Systems, Moscow, Russia}, 2017.

\bibitem[]{Furfaro:2016:Robust_ZEMZEV} Furfaro, R. and Wibben, R. D. , ``Robustification of a class of guidance algorithms for planetary landing: Theory and applications'' \emph{26th AAS/AIAA Space Flight Mechanics Meeting, 2016. Univelt Inc.}, 2016.


\bibitem[]{Huang:2015:ELM} Huang, G. B. , ``What are extreme learning machines? Filling the gap between Frank Rosenblatt’s dream and John von Neumann’s puzzle'' \emph{Cognitive Computation 7.3}, 2015.

\bibitem[]{Huang:2011:ELM_survey} Huang, G. B. and Wang, D. H. and Lan, Y. , ``Extreme learning machines: a survey'' \emph{International journal of machine learning and cybernetics}, Vol. 2, No. 2, pp. 107-122, 2011.


\bibitem[]{Silver:2014:deterministic_PG} Silver, D. and Lever, G. and Heess, N. and Degris, T. and Wierstra, D. and Riedmiller, M. , ``Deterministic Policy Gradient Algorithms'' \emph{ICML}, 2014.


\bibitem[]{Sutton:1998:sutton_reinforcement_learning} Sutton, R. S. and Barto, A. G. , ``Reinforcement learning: An introduction'' \emph{Cambridge: MIT press}, Vol. 1, No. 1, 1998.

\bibitem[]{Sutton:2000:PG_convergence} Sutton, R. S. and McAllester, D. and Singh, S. and Mansour, Y. , ``Policy gradient methods for reinforcement learning with function approximation'' \emph{Advances in neural information processing systems}, 2000.


\bibitem[]{Williams:1992:sutton_reinforcement_learning}Williams, R. J. , ``Reinforcement learning'' \emph{Springer}, 1992.


\bibitem[]{Wu:1978:stability_wu} Wu, M. , ``Transformation of a linear time-varying system into a linear time-invariant system'' \emph{International Journal of Control}, Vol. 24, No. 4, pp. 589-602, 1978.

\bibitem[]{Hagan:1994:backProp} Hagan, M. T. and Menhaj, M. B. , ``Training feedforward networks with the Marquardt algorithm'' \emph{IEEE transactions on Neural Networks}, Vol. 5, No. 6, pp. 989-993, 1994.

\bibitem[]{Rowell:2002:STM} Rowell, D. , ``Time-domain solution of LTI state equations'' \emph{Class Handout in Analysis and Design of Feedback Control System}, 2002.

\bibitem[]{Patterson:2014:GPOPS} Patterson, Michael A., and Anil V. Rao. ``GPOPS-II: A MATLAB software for solving multiple-phase optimal control problems using hp-adaptive Gaussian quadrature collocation methods and sparse nonlinear programming`` ACM Transactions on Mathematical Software (TOMS) 41.1 (2014): 1-37.


\end{thebibliography}

\end{document}